\documentclass[aip,twocolumn,amsmath,amssymb,hyperref,reprint]{revtex4-1}

\usepackage{tabularx}
\usepackage{graphicx}
\usepackage{array}
\usepackage{amssymb}
\usepackage{amsfonts}
\usepackage{amsmath}
\usepackage{epstopdf}
\usepackage{mathrsfs}
\usepackage{color}
\usepackage{xcolor}
\usepackage{booktabs}
\usepackage{multirow}
\usepackage{subfigure}
\usepackage{epsfig}
\usepackage{chngpage}
\usepackage{latexsym}
\usepackage{tabu}
\usepackage{threeparttable}

\begin{document}

\title{A model for meme popularity growth in social networking systems based on biological principle and human interest dynamics}

\date\today

\author{Le-Zhi Wang}
\affiliation{School of Electrical, Computer and Energy Engineering, Arizona State University, Tempe, Arizona 85287, USA}

\author{Zhi-Dan Zhao}
\affiliation{School of Electrical, Computer and Energy Engineering, Arizona State University, Tempe, Arizona 85287, USA}
\affiliation{Web Sciences Center, Big Data Research Center, University of Electronic Science and Technology of China, Chengdu 610054, China.}

\author{Jun-Jie Jiang}
\affiliation{School of Electrical, Computer and Energy Engineering, Arizona State University, Tempe, Arizona 85287, USA}

\author{Bing-Hui Guo}
\affiliation{School of Mathematics, Beihang University, Beijing 100191, China}

\author{Xiao Wang}
\affiliation{School of Biological and Health Systems Engineering, Arizona State University, Tempe, AZ 85287, USA}

\author{Zi-Gang Huang} \email{huangzg@xjtu.edu.cn}
\affiliation{School of Life Science and Technology, Xi'an Jiaotong University, Xi'an 710049, China}
\affiliation{Institute of Computational Physics and Complex Systems, Lanzhou University, Lanzhou, Gansu 730000, China}

\author{Ying-Cheng Lai} \email{Ying-Cheng.Lai@asu.edu}
\affiliation{School of Electrical, Computer and Energy Engineering, Arizona State University, Tempe, Arizona 85287, USA}
\affiliation{Department of Physics, Arizona State University, Tempe, Arizona 85287, USA}

\begin{abstract}

We analyze five big data sets from a variety of online social networking (OSN) systems and find that the growth dynamics of meme popularity exhibit characteristically different behaviors. For example, there is linear growth associated with online recommendation and sharing platforms, a plateaued (or an ``S''-shape) type of growth behavior in a web service devoted to helping users to collect bookmarks, and an exponential increase on the largest and most popular microblogging website in China. Does a universal mechanism with a common set of dynamical rules exist, which can explain these empirically observed, distinct growth behaviors? We provide an affirmative answer in this paper. In particular, inspired by biomimicry to take advantage of cell population growth dynamics in microbial ecology, we construct a base growth model for meme popularity in OSNs. We then take into account human factors by incorporating a general model of human interest dynamics into the base model. The final hybrid model contains a small number of free parameters that can be estimated purely from data. We demonstrate that our model is universal in the sense that, with a few parameters estimated from data, it can successfully predict the distinct meme growth dynamics. Our study represents a successful effort to exploit principles in biology to understand online social behaviors by incorporating the traditional microbial growth model into meme popularity. Our model can be used to gain insights into critical issues such as classification, robustness, optimization, and control of OSN systems.

\end{abstract}
\maketitle

{\bf
With advances in information technologies, a novel class of complex dynamical
systems has emerged: online social networking (OSN) systems. The complexity
of OSN systems is enormous: posting and sharing of messages by users, sudden 
occurrence of breaking news events, and random drifts in user interests, 
etc., all leading to drastic variations of the network structure and dynamics 
with time and making (big) data analysis an essential approach to uncovering 
the inner dynamical working of these systems. A phenomenon that has attracted 
recent attention is growth dynamics of memes such as news, ideas, knowledge 
or rumors in OSN systems. Previous models focusing on the individual level 
were unable to account for the common phenomenon of group popularity, raising 
the need to develop a comprehensive model that incorporates heterogeneity of 
users and memes to describe quantitatively the collective dynamics of meme 
popularity. Another challenge in the construction of a model for meme 
popularity lies in its distinct growth behaviors in different OSN systems. 
Our analysis of five big data sets from diverse social networking platforms
has revealed at least three characteristically different types of behaviors: 
linear, plateaued (or ``S'' shaped), and exponential growth in time. Is it 
possible to construct a single model that can explain the distinct growth 
behaviors? This paper provides an affirmative answer. The general principle
underlying our work is that, while OSN systems are human-engineered with 
vast complexity, nature has solved difficult problems in complex systems. 
Animals, plants, microbes, and even cells are extremely well self-organized 
natural systems with superior functions and efficiency. The first ingredient
of our model is then an approximate equivalence between meme evolution in OSN 
systems and microbial cell population growth. This leads to a probabilistic, 
population-level base model, where at any given time a cell can experience 
one of the three possible events: division (generation), death, and survival. 
Regarding memes as the ``microscopic'' elements of OSN systems, 
the possible events that can happen to a meme are similar: 
posting/forwarding, being overwritten (exclusion), or simple survival, which 
are equivalent, respectively, to cell division, death, and survival in a 
microbial system. Because meme growth is a kind of human behavior, 
it is also necessary consider additional model ingredients beyond the 
biological equivalence. The second ingredient of our model is then to 
incorporate human-interest dynamics into the bio-inspired base growth model. 
The outcome is a hybrid model for meme popularity dynamics, which contains 
four free parameters that can be determined from data. The striking result 
is that the model can predict the detailed meme popularity growth behaviors 
in all real OSN systems studied, regardless of their characteristically 
distinct origins, thereby providing a solid ground for its validity and 
universal applicability. To be able to predict the dynamical evolution of 
memes is of great social, economic, and political interest. What we have 
achieved in this paper is a universal model for this task with minimally 
required information.
}

\section{Introduction} \label{sec:intro}

Online social networking (OSN) systems are now ubiquitous and play
an increasingly important role in the modern society, as they provide
unprecedented platforms supporting communications among a vast number
of users all over the world. Due to the availability of massive 
data sets from OSN systems, quantitative system analyses become
possible~\cite{HN:2006,SBB:2010,CHBG:2010,OR:2010,RMK:2011,YL:2011,
GPV:2011,WFVM:2012,GWG:2012,BMLB:2012,GAHY:2012,YAH:2012,OI:2013,
Zhaoetal:2013, WHRWL:2014,GWOL:2014,LKML:2014,GCOPR:2014,KB:2015,GOBM:2016,Zhangetal:2016,WWZJ:2016,QOSFM:2017,LMSTM:2017}. Previous efforts focused on 
issues such as network and opinion co-evolution~\cite{HN:2006},
user behavior modeling on networks~\cite{SBB:2010,GPV:2011}, the dynamics of
users' activity across topics and time~\cite{CHBG:2010,BMLB:2012}, human
interest dynamics in e-commerce and communication~\cite{Zhaoetal:2013},
evolutionary dynamics of forwarding network in the Weibo 
platform~\cite{WHRWL:2014}, competition among different Twitter
topics~\cite{WFVM:2012,LGR:2012,GWOL:2014,GOBM:2016}, popular topic-style 
analyses in the Twitter-like social media~\cite{GAHY:2012,YAH:2012,OI:2013},
information diffusion patterns in different domains~\cite{RMK:2011,GWG:2012},
and the effect of coexistence of different OSN network services~\cite{KB:2015}.
These studies mainly considered two issues common to various OSN systems:
the collective behaviors of users and the dynamics of posts or memes.
Based on empirical findings, e.g., power-law scaling relations, mathematical 
and/or physical models have developed to predict the scaling laws. For example,
a two-layer model has been proposed to characterize the viral dynamics and 
media influence~\cite{KB:2014}, a branching process has been used to explain 
the power-law distribution of meme popularity~\cite{GWOL:2014,GOBM:2016}, and 
a Bayesian probabilistic model has been developed to characterize the 
evolution of tweets~\cite{ZFB:2014}. While these models are able to simulate 
or predict certain aspects of meme popularity in real OSN systems, they are 
often dependent upon the structure of the underlying social network, limiting 
their applicability to specific types of social networking platforms with 
specialized functions. Due to the vast complexity and diversity of the OSN 
systems, a quantitative, generally applicable model for the dynamical 
evolution of key variables of empirical interest is lacking. In this paper, 
exploiting biological principles, we develop a universal model to explain 
the characteristically distinct behaviors of meme growth observed from 
diverse OSN systems. 

Previous efforts in this field are briefly summarized, as follows. We define 
memes broadly as some items that serve to attract user attention and induce 
heterogeneous dynamical behaviors in the OSN. Especially, memes are referred 
to not only as news pieces, ideas, certain information pieces, knowledge 
items or rumors, but also as bookmarks, movies, Weibo messages, and music 
pieces, etc. The network to characterize the user-item relationship typically 
possesses a bipartite structure~\cite{BHBGLM:2017}. Some memes can go viral, 
some might receive constant attention, and some simply get ignored. To uncover 
the mechanisms that drive the fates of different types of memes on a 
microscopic scale is a challenging task. In this regard, in a previous 
work~\cite{WFVM:2012}, it was found that in OSN systems, the distribution of 
meme popularity is typically heterogeneous as a result of the mutual 
``competition'' among different coexisting memes for users' attention. 
This observation provides the base for the proposal of a
theoretical model to describe the dynamical evolution of memes with
a particular focus on the influence of user actions on information
diffusion~\cite{GWOL:2014,GOBM:2016}. The mathematical backbone of the
theory is branching processes and it has successfully explained certain
empirical observations such as the distribution of meme popularity growth
associated with the {\em Twitter} data sets. A key assumption of the theory,
which is somewhat ideal and thus makes feasible an analytic treatment, is that
users have constant activity rates and memes are equipped with the same
``fitness.'' While the idealization of identical users and memes enables 
a mathematical analysis, the key ingredient of meme dynamics in real world 
OSN systems is heterogeneity in user and meme behaviors. An alternative 
modeling approach was to apply a self-exciting point process (e.g., the Hawkes 
process~\cite{Hawkes:1971}) to predict the popularity of tweets based on 
partial information about the network structure and observations of the 
retweeting times~\cite{ZFB:2014,ZEHRL:2015}. This type of models can 
successfully predict the total final number of retweets but, because of the 
focus on information diffusion at the individual level, they are unable to 
account for the common phenomenon of group popularity. It is also worth 
noting that memes popularity dynamics are distinct from epidemic spreading 
dynamics on complex networks~\cite{PSRV:2001,Newman:2002,PCVV:2015} in that, 
in the former, individuals receive and spread a large number of memes while 
in the latter, the type of viruses is typically one or two. The theories 
and computational methods developed in the past on network spreading 
dynamics~\cite{PCVV:2015} are generally {\em not} applicable to meme 
popularity dynamics.

At the present, a comprehensive model that incorporates the heterogeneity of 
users and memes to describe quantitatively the collective dynamics of meme 
popularity is lacking. A more significant challenge in the construction of 
a model for meme popularity lies in its distinct growth behaviors in different 
OSN systems. In particular, by analyzing five big data sets from diverse 
social networking platforms, we find three characteristically distinct types 
of behaviors: linear, plateaued (or ``S'' shaped), and exponential growth in 
time. Is it possible to construct a single model that can explain these 
distinct growth behaviors? Naturally, such a model will contain a small number
of free parameters whose values depend on the specific OSN system and
can be estimated from data. Except for the differences in the values of
the free parameters, the basic elements of the model are identical for
OSN systems from diverse contexts. In this sense the model can be regarded
as universal. Our main idea is to exploit biological principles (biomimicry) 
to develop such a model. The guiding principle is that, while OSN systems are 
man-made with vast complexity, nature has solved difficult problems in more 
complex systems, especially in biology. Nevertheless, because meme growth
dynamics are driven by human behaviors, it is also necessary to incorporate 
human aspects into the model. For this purpose we exploit a model for human
interest dynamics~\cite{Zhaoetal:2013}. The final outcome is a hybrid model 
for meme growth based on the combination of biomimicry inspired by cell 
growth in microbial ecology, human behaviors, and empirical laws extracted 
from big data sets. The model can accurately predict characteristically 
distinct growth behaviors in a diverse array of OSN systems.

In Sec.~\ref{sec:biomimicry}, we identify the similarities between cell growth 
in microbial ecology and meme growth in OSNs to establish our biomimicry 
principle. In Sec.~\ref{sec:model}, we develop a hybrid model incorporating
human interest dynamics into our cell-growth model for meme evolution. In 
Sec.~\ref{sec:results}, we present numerical support for our hybrid model 
based on five big data sets from OSN systems and provide a mathematical 
analysis. In Sec.~\ref{sec:discussion}, we summarize the main results and 
discuss possible model generalization.
 
\section{Biomimicry principle and empirical support}
\label{sec:biomimicry}
 
\subsection{Biomimicry: from microbial ecology to social network}

In microbial ecology, quantitative analysis of cell population is
fundamental~\cite{von:1959,Trucco1:1965,Trucco2:1965,LVTH:2004,
KRB:2005,Wangetal:2010,Horowitz:2010,SAKWS:2013,Ribeiro:2014,AS:2015,
GC:2016,CG:2016,Widder:2016}. Due to computational constraints, most cell
models are at the population level because it is impractical to monitor the
state of each individual cell and count the number of living cells at every
time step. A typical cell population model contains three ingredients: cell 
division (birth), death, and survival. Likewise, in an OSN system, it is 
unrealistic to study the behavior of every single post, but it is feasible to 
obtain data reflecting the collective behaviors of these posts. Especially,
a meme can be generated from one user and passed onto another in response to
certain social event (birth), it can disappear if there is no or no longer
any interest in it (death), or it can simply be associated with the same user
without any change (survival). This analog suggests that, dynamically, a
meme associated with an individual is equivalent to a cell in a microbial
system.

\begin{figure*}
\centering
\includegraphics[width=\linewidth]{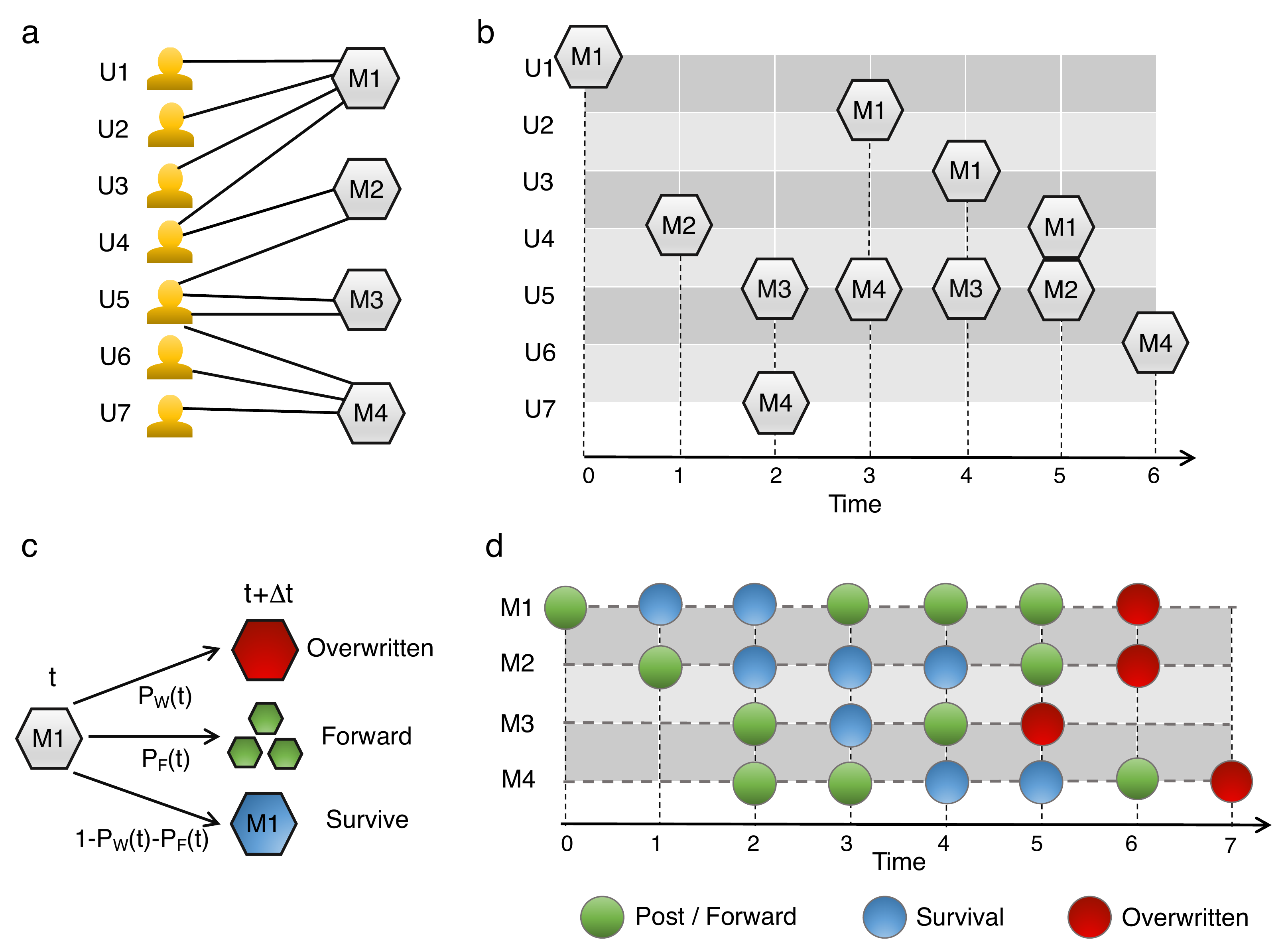}
\caption{ {\bf Schematic illustration of evolutionary dynamics of meme
popularity}. The toy system has seven users and four types of memes, where a
meme associated with an individual user corresponds to a cell in microbial
ecology. (a) Bipartite graph representation of users and memes, which are
connected with each other by posting or forwarding actions. (b) Representation 
of forwarding actions at different time. Each horizontal line is associated 
with an individual user and each hexagon represents a meme. For a horizontal 
line, a hexagon appearing at a time indicates that a meme (regardless of its 
type) has been created or forwarded at this time. (c) The three events that can
occur to a meme at a given time, together with the respective probabilities:
$P_F(t)$ - the probability of being forwarded, $P_W(t)$ - the probability
of being overwritten (exclusion probability), and the survival probability
$1 - P_F(t) - P_W(t)$. (d) An example of the time evolution of meme
popularity, where the green circles represent posting or forwarding
events, the blue circles define survival events, and the red circles
correspond to exclusion events. A meme is regarded as dead (excluded)
after the last forwarding event.}
\label{fig:illustration}
\end{figure*}

To place the cell-meme correspondence on a quantitative footing, we consider
the probabilistic model of cell growth and mortality~\cite{Horowitz:2010}.
At each basic time step, three possible events can occur to an individual
cell: it can divide, can die, or remain alive. The probabilities of the
respective events are represented as functions of time with parameters
estimated from experimental data. Considerations of the events and the
associated probabilities provide a base for us to formulate a meme 
probabilistic model. An illustrative example is presented in 
Fig.~\ref{fig:illustration}. In this toy OSN system, there are seven users 
and four types of memes that constitute a bipartite network, as shown in 
Fig.~\ref{fig:illustration}(a), where the former and the latter are connected
with each other by posting or forwarding actions. Especially, two users are 
directionally connected if a meme is forwarded from one to another. 
Figure~\ref{fig:illustration}(b) shows an example of the posting or 
forwarding action for each type of meme from the users point of view, 
which is essentially a space-time representation of how the memes are 
created and evolve. 
Here, $t = 0$ specifies the initial time of the observational window of the 
system. For example, users $U1$, $U4$, $U5$ and $U7$ post meme types $M1$,
$M2$, $M3$ and $M4$ at time $t=0$, $t=1$, $t=2$, $t=2$, respectively, while
users $U2$, $U3$ and $U4$ forward the $M1$ type at $t=3$, $t=4$ and $t=5$,
respectively. As a result, four users ($U1$, $U2$, $U3$, and $U4$) are linked
to the meme type $M1$, as indicated in the bipartite network in
Fig.~\ref{fig:illustration}(a). Similarly, user $U5$ forwards meme type $M4$
at $t=3$, $M3$ at $t=4$ and $M2$ at $t=5$, and user $U6$ forwards type $M4$
at $t=6$. For a meme associated with a user at a given time, it can be
forwarded, be overwritten (excluded), or simply survive, as indicated in
Fig.~\ref{fig:illustration}(c), which corresponds to the three possible events
that can occur to a cell in microbial ecology: division, death, or remaining
alive. Figure~\ref{fig:illustration}(d) shows the same example as in 
Fig.~\ref{fig:illustration}(b), but from meme's point of view to illustrate 
the three events for each of the four meme types. 
From a global perspective, a population of a specific meme starts to grow 
when it is first posted. The population size increases as users continue
to forward this meme. Between each pair of consecutive forwarding events, 
the meme is in the survival state. After the last forwarding event, the 
meme is regarded as being excluded or dead.

The population of a type of memes at any time is determined by the numbers
of newly forwarded, excluded, and survived posts. Let $F(t)$, $W(t)$,
and $S(t)$ be the numbers of forwarded, excluded, and survived memes,
respectively, at time $t$. The meme population at this time can be
written as
\begin{equation} \label{eq:Nt}
N(t)=S(t)+F(t).
\end{equation}
The fraction of meme population is defined as 
\begin{displaymath}
{{\rm{P}}_N}(t) = ({{\frac{{S(t) + W(t)}}{{F(t) + S(t) + W(t)}})} 
\mathord{\left/{\vphantom {{\frac{{S(t) + W(t)}}{{F(t) + S(t) + W(t)}})} 
{\max (\frac{{S(t) + W(t)}}{{F(t) + S(t) + W(t)}})}}} \right.
\kern-\nulldelimiterspace} {\max (\frac{{S(t)+W(t)}}{{F(t)+S(t)+W(t)}})}},
\end{displaymath}
and the interactions among memes are described by the following normalized 
forwarding and exclusion probability functions: 
\begin{eqnarray}
\nonumber
{{\rm{P}}_F}(t) & = & ({{\frac{{F(t)}}{{F(t) + S(t) + W(t)}})} \mathord{\left/
 {\vphantom {{\frac{{F(t)}}{{F(t) + S(t) + W(t)}})} {\max (\frac{{F(t)}}{{F(t) + S(t) + W(t)}})}}} \right. \kern-\nulldelimiterspace} 
{\max (\frac{{F(t)}}{{F(t) + S(t) + W(t)}})}}, \\ \nonumber
{{\rm{P}}_W}(t) & = & ({{\frac{{W(t)}}{{F(t) + S(t) + W(t)}})} \mathord{\left/
 {\vphantom {{\frac{{W(t)}}{{F(t) + S(t) + W(t)}})} {\max (\frac{{W(t)}}{{F(t) 
+ S(t) + W(t)}})}}} \right. \kern-\nulldelimiterspace} 
{\max (\frac{{W(t)}}{{F(t) + S(t) + W(t)}})}}, 
\end{eqnarray}
respectively, which determine the values of $F(t)$ and $W(t)$.

\subsection{Validation of biomimicry principle with empirical online
data sets}

\begin{table*}
\caption{Basic properties of four data sets studied in this paper}
\begin{center}
\begin{tabularx}{1\textwidth}{p{0.18\textwidth}p{0.15\textwidth}p{0.15\textwidth}p{0.15\textwidth}p{0.15\textwidth}p{0.15\textwidth}}
\hline
Data Sets & Records & Memes & Users & Duration (Months) & References \\
 \hline
	Delicious &  361,928,091 & 886,405 & 43,968,955 & 42   & [\onlinecite{WZB:2008}] \\
	Douban Book & 20,199,759 &  455,177 & 557,879  & 53  & [\onlinecite{Zhaoetal:2013}] \\
	Douban Movie & 65,205,220 &  504,066 &  86,503 & 52  & [\onlinecite{Zhaoetal:2013}] \\
	Douban Music & 25,596,271 &  403,835 &  395,035 & 50 & [\onlinecite{Zhaoetal:2013}] \\
\hline
\end{tabularx}
\end{center}
\begin{tablenotes}[flushleft]
\small
\item *The resolution unit of time is Days.
\end{tablenotes}
\label{table:datasets}
\end{table*}

The massive empirical data sets analyzed in this article are from
large-scale online systems: {\em Delicious}, {\em Douban} and {\em Weibo}.
The basic statistical properties of four data sets are listed in
Table~\ref{table:datasets}, where the term ``Records'' represents the
number of records in each raw data set, ``Memes'' denotes the total
number of memes in each raw data set, ``Users'' indicates the total
number of users involved in each raw data set, and ``Duration'' is
the duration of each processed data set. The details of these
data sets are described, as follows.

{\em Delicious} is a web service focusing on helping users collect
bookmarks. Each record consists of the operation time, user's ID, Universal
Resource Locator (URL) and the tag of URL. In this data set, a meme is
defined as a bookmark, and users' collections of bookmarks are regarded as
forwarding actions. An ``excluded'' (or ``overwritten'') bookmark at time 
$t$ means that this bookmark no longer appears in the system after time $t$.

{\em Douban} is a major Chinese Social Networking Service website. It
allows users to record information and make recommendations related to
books, movies, and music, etc. Each record contains users id, time stamps
and item rating actions. We analyze three Douban
data sets: {\em Douban Book}, {\em Douban Movie} and {\em Douban Music}.
We define each rated item as a meme, and treat each rating action as a
forwarding event. A book (movie or a piece of music) not recurring at 
time $t$ is regarded as ``excluded'' (or ``overwritten'').

{\em Sina Weibo} is by far the largest and most popular microblogging
website in China: it is a widely used twitter-like microblogging social
network medium with more than $500$ million registered users in
China~\cite{weibo:2015}. The appealing features of the data include
wide publicity, real-time availability of information, and message
compactness. Similar to Twitter, Weibo attracts users through all kinds
of breaking news and spotlight topics. All users can see messages, called
Weibo in Chinese, published by concerned users. Given a specific topic
of interest, an individual can participate by retweeting (forwarding)
or tweeting (posting) any interesting Weibo~\cite{BGL:2010}. In this data
set, for each message with the forwarding information, we record the original
Weibo id, user id who forwards this Weibo and the time of creation. Each
message represents a meme with a possible forwarding action.

\begin{figure*}[htb]
\centering
\includegraphics[width=1\linewidth]{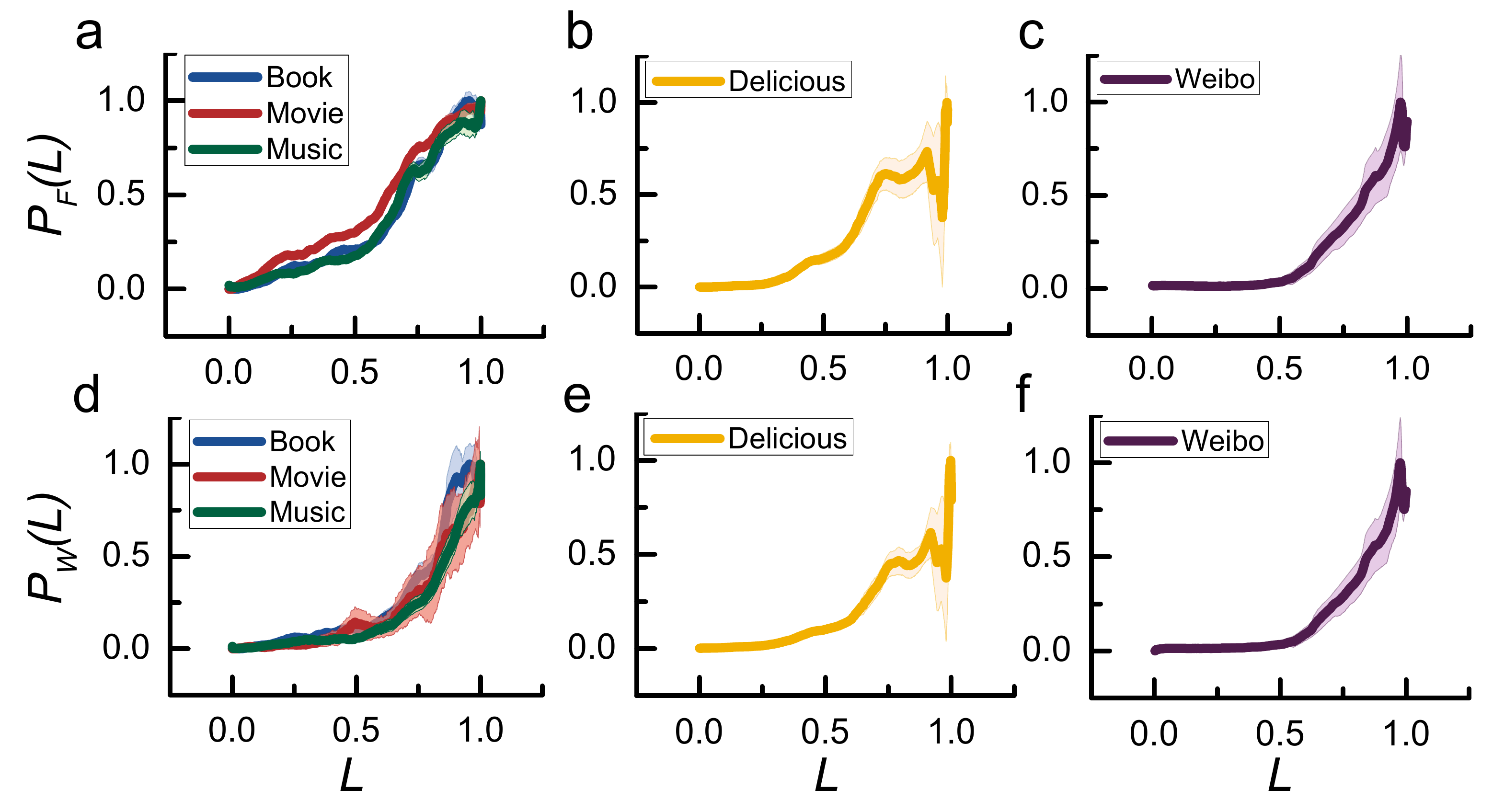}
\caption{ {\bf Time evolution of forwarding and exclusion probabilities
estimated from empirical data sets}. In each panel, lines and the
corresponding shadow areas indicate the average forwarding or exclusion
probabilities and the error bars, respectively. The time and the probabilities
have been normalized to facilitate a quantitative comparison among the
five data sets that arise from different social networking contexts. The
probabilities are for the group of data sets {\em Douban~Book, Movie} and
{\em Music} (a,d), {\em Delicious} (b,e), and {\em Weibo} (c,f).
Note that, it is not necessary that the probabilities $P_A(L)$ and $P_W(L)$ 
are increasing functions of time. However, since the focus is on the meme 
growth dynamics, in the normalized time interval considered ($L \in [0,1]$),
these probabilities exhibit an increasing trend. Also, since the probabilities 
are normalized by their respective maximum values at $L = 1$, their values 
being unity does not mean that all users are active at $L = 1$.}
\label{fig:realdata}
\end{figure*}

The probabilities $P_F(t)$ and $P_W(t)$ are key quantities in our meme
popularity model, which can be estimated from the empirical data.
Figure~\ref{fig:realdata} shows the forwarding and exclusion probabilities 
for the five data sets: {\em Delicious}, {\em Douban Book}, {\em Douban Movie},
{\em Douban Music}, and {\em Sina Weibo} (or {\em Weibo}). We find it 
convenient to organize the five data sets into three groups in terms of their 
nature: {\em Douban Book, Movie and Music} (group 1), 
{\em Delicious} (group 2), and {\em Weibo} (group 3). 
To facilitate a comparison of the dynamical behaviors of different systems, 
we define the normalized time $t \to L$. For convenience, we choose the 
range of $L$ to be $[0,1]$, where $L = 1$ is defined as the time when the 
fraction $P_N(L)$ of meme population reaches maximum. The probabilities
${{\rm{P}}_F}(t) \to {{\rm{P}}_F}(L)$ and ${{\rm{P}}_W}(t) \to {{\rm{P}}_W}(L)$
are also normalized by their respective maximum values at $L = 1$. 
Initially, the two normalized probabilities have
relatively low values but they begin to increase after certain time. This
initial ``silent'' phase corresponds to the lag phase that occurs before the
accumulation phase for cell growth in microbial ecology~\cite{von:1959,
Trucco1:1965,Trucco2:1965,LVTH:2004,KRB:2005,Wangetal:2010,Horowitz:2010,
SAKWS:2013,Ribeiro:2014,AS:2015,GC:2016,CG:2016,Widder:2016},
where the growth and mortality rates are expected to be low at the beginning
but increase with time. The striking phenomenon is that, for the five data
sets arising from diverse social networking contexts, the time evolution
of the probabilities exhibits quite similar features, suggesting a universal
mechanism underlying the dynamical evolution of meme popularity.

\section{Model construction} \label{sec:model}

\subsection{Basic principles underlying the construction of a 
universal model for meme popularity dynamics} 

\begin{figure*}
\centering
\includegraphics[width=\linewidth]{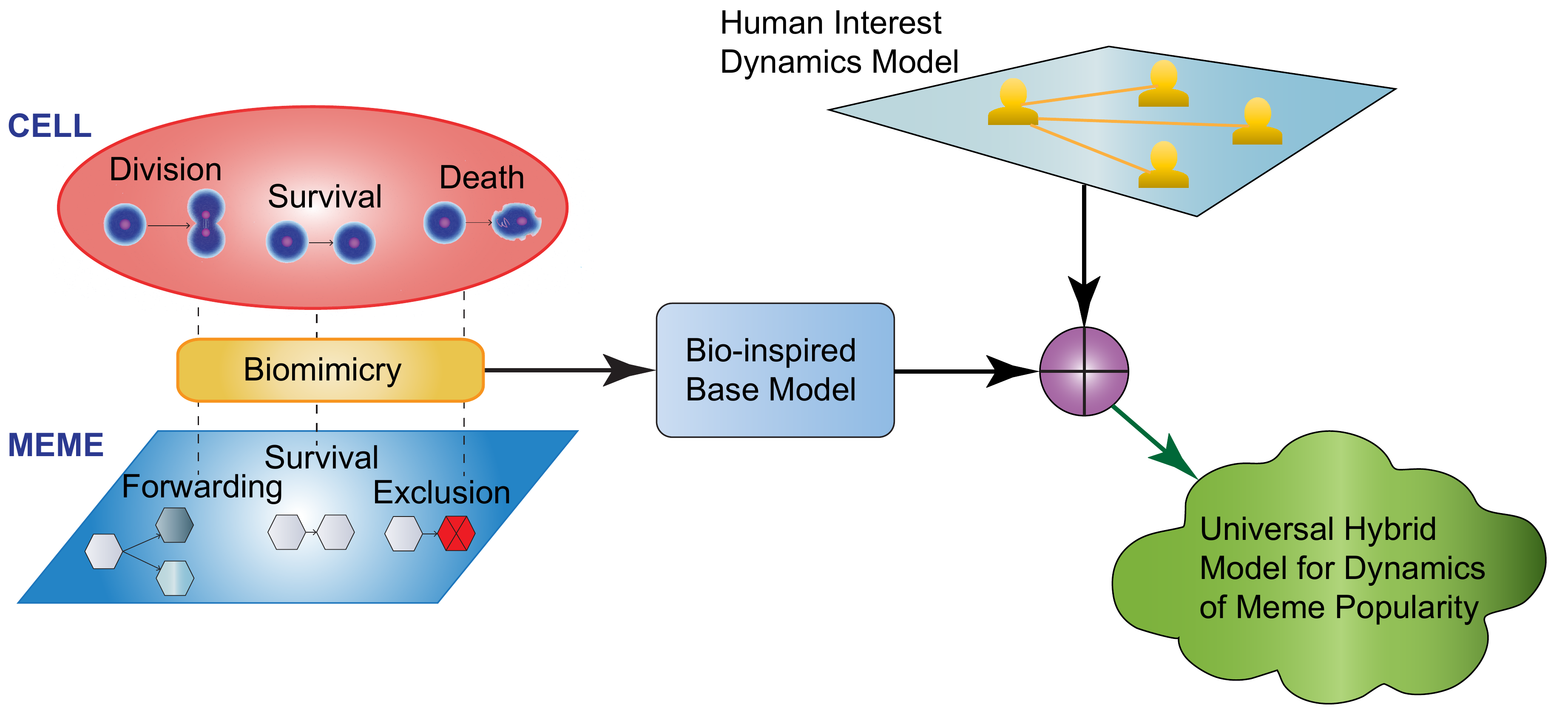}
\caption{ {\bf Illustration of the basic principles underlying the
construction of a universal model for meme popularity dynamics: biomimicry
and human interest dynamics}. Cell population dynamics in microbial ecology
contain three basic elements: cell division, death, and survival. The
dynamical evolution of memes in OSN systems has three corresponding
elements: forwarding, exclusion, and survival. A base model for meme
popularity dynamics can then be constructed according to the cell growth
model in biology. Human factors, however, play a significant role in
meme evolution and therefore must be taken into account.
A combination of the biology inspired base model and human interest
dynamics model leads to a universal hybrid model for meme popularity
dynamics.}
\label{fig:scheme}
\end{figure*}

Our first step is to hypothesize the equivalence between meme evolution in 
OSN systems and microbial cell population growth so as to develop a 
probabilistic, population-level base model. In such a dynamical evolution 
model of cell population~\cite{von:1959,Trucco1:1965,
Trucco2:1965,LVTH:2004,KRB:2005,Wangetal:2010,Horowitz:2010,SAKWS:2013,
Ribeiro:2014,AS:2015,GC:2016,CG:2016,Widder:2016}, at any given time a cell
can experience one of the three possible events: division (generation),
death, and survival. Likewise, memes are the ``microscopic'' elements of
OSN systems. At a given time, the possible events that can happen to
a meme are similar: posting/forwarding, being overwritten (exclusion),
or simple survival, which are equivalent, respectively, to cell division,
death, and survival in a microbial system, as schematically illustrated on
the left side of Fig.~\ref{fig:scheme}. 
This equivalence, or biomimicry, leads to a base model for meme population 
growth. Additional model ingredients beyond the biological
equivalence must be sought. This is reasonable as OSN systems are man-made
and, as such, human factors can play a significant role in the dynamics.
The second step in our model building is then to incorporate human interest
dynamics that, intuitively, are correlated with meme
popularity. Incorporating a general model for the dynamical evolution of
human interest into the bio-inspired base growth model, we arrive at a
hybrid model for meme popularity dynamics, as shown on the right side of
Fig.~\ref{fig:scheme}. The final model contains four free parameters that
can be determined from data, as we demonstrate using empirical big data
sets from diverse OSN systems in interest sharing (e.g., various {\em Douban}
platforms and {\em Delicious}) and the online OSN platform {\em Weibo}.
The striking result is that the model can predict the detailed meme popularity
growth behaviors in all real OSN systems studied, regardless of their
characteristically distinct origins, thereby providing a solid ground
for its validity and universal applicability. While our model predicts
three distinct meme popularity growth behaviors from five data sets, with 
the values of the four free parameters determined from data, the model
has the capability to predict growth behaviors beyond the three types.

It should be emphasized that, while the cell growth model captures certain 
features of the meme dynamics, our data analysis indicates that large errors 
can arise when attempting to predict the detailed growth dynamics of meme 
population in terms of the time evolution of the forwarding and exclusion 
probabilities estimated directly from the empirical data, as shown in 
Fig.~\ref{fig:realdata}. Especially, while the biologically 
motivated model is able to provide an overall picture (statistical behaviors) 
about the meme growth dynamics, it is not possible for it to capture the 
minuscule details in each data set. 

\subsection{Beyond biomimicry: incorporation of human interest dynamics} 

\begin{figure*}[htb]
\centering
\includegraphics*[width=1\linewidth]{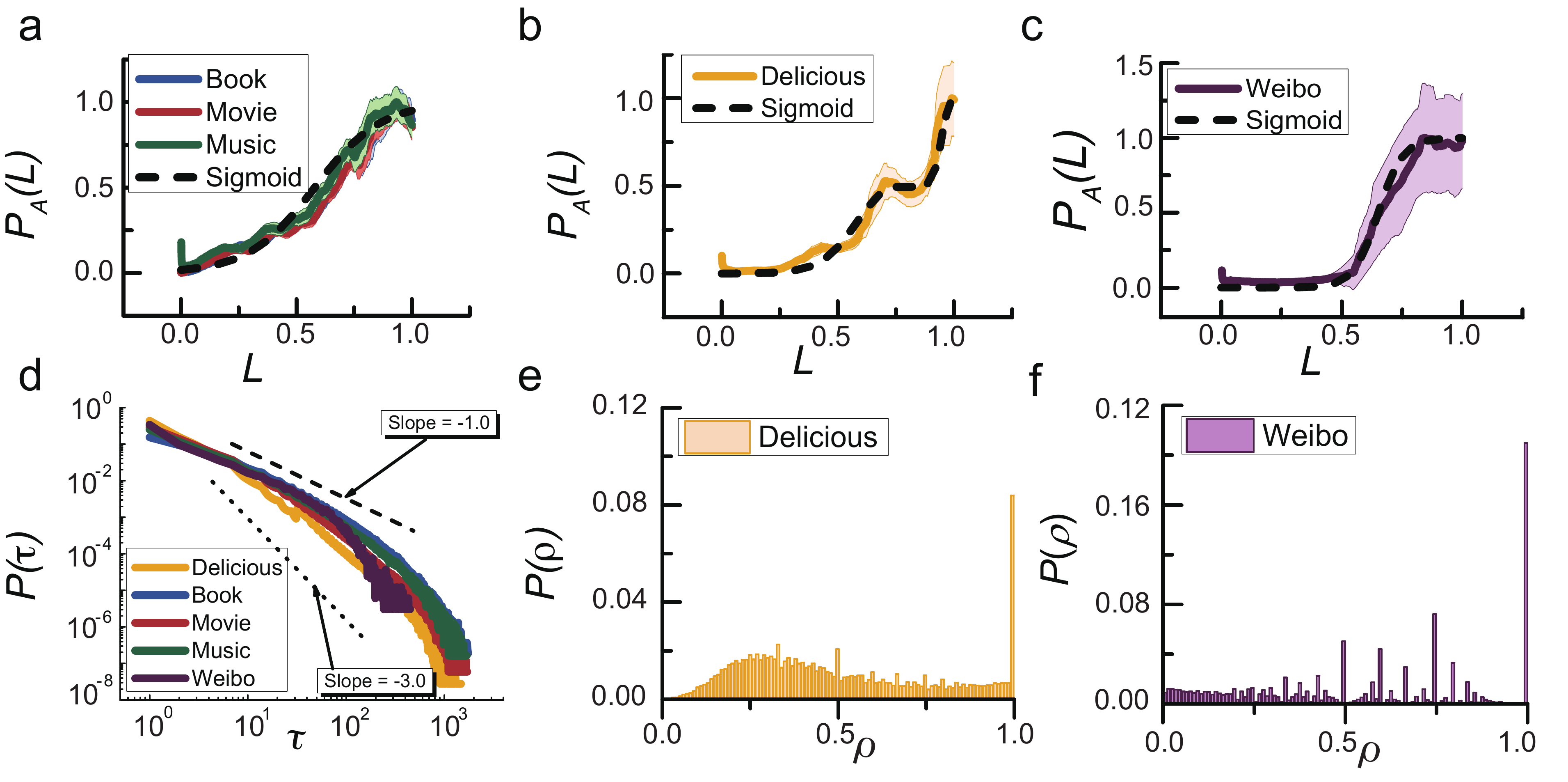}
\caption{ {\bf Users behaviors extracted from the empirical data sets with
respect to human interest dynamics}. (a-c) Activation curves (the
fraction of users activated at normalized time $L$) together with the error
bars for the three groups of empirical data, where the yellow, blue, red,
green and purple solid lines and the corresponding shadow areas represent
the average activation rates and error bars for {\em Delicious},
{\em Douban~Book}, {\em Douban~Movie}, {\em Douban~Music}, and {\em Weibo},
respectively. The black dash curves are the fitted sigmoid functions.
For (a), the user activation curves are from the three data sets:
{\em Douban~Book}, {\em Movie} and {\em Music}, which exhibit a striking
agreement at a detailed level, and can be fitted by a single sigmoid function.
For the data set {\em Delicious} in (b), a plateau in the user activation
curve emerges, which can be fitted by two distinct sigmoid functions.
For the data set {\em Weibo} in (c), the activation curve can be fitted by
a single sigmoid function. (d) Algebraic distribution of the individual
interevent time for the five data sets, where the algebraic scaling exponent
$\alpha$ has values ranging from one to three. (e,f) Distributions of
parameter $\rho$ from the data sets {\em Delicious} and {\em Weibo},
respectively. See Table~\ref{table:fitreal} for the fitting values of
all model parameters from data, which are estimated from least-squares 
fitting subject to the standard Kolmogorov-Smirnov (KS) test~\cite{CF:2009} 
with $D = 0.1$.}
\label{fig:user_real}
\end{figure*}

While biomimicry can provide pivotal insights into uncovering
the laws governing human social networking systems, certain aspects of the
human behaviors are fundamentally absent in any biological system. In
particular, in the context of our present study, while microbial cell
evolution provides a base for constructing models for online social behaviors, 
it has a purely biological origin while OSN systems are man-made. There ought 
to be some differences between the two types of systems. Indeed, in spite of 
the remarkable agreement among the forwarding and exclusion probability
functions across OSN systems of different nature (Fig.~\ref{fig:realdata}),
a detailed examination of the evolution of the meme population $N(t)$ for
the three groups of data sets reveals some discrepancies, indicating the
need to include additional factors that are not present in the microbial
cell evolution model. In fact, it is necessary to incorporate human factors
into models of OSN systems. For the dynamical evolution of meme popularity,
the most pertinent factor is human interest.

Meme growth in OSN systems is a human behavior. In general, human behaviors
are driven by human interest. In the past decade or so, there has been a
great deal of effort in modeling and understanding human behavior and interest
dynamics~\cite{Barabasi:2005,OB:2005,Dezso:2006,ZKKWH:2008,GR:2008,SKWB:2010,
SSPTL:2012,Zhaoetal:2013}. Of particular interest are characteristics such as
the distribution of the interevent time of human behaviors, the distribution
of the return time to revisit a particular interest, interest ranking and
transition, and the distribution of the time that an interest lasts.
Because of the sensitive dependence of human interest on environment factors,
it was previously speculated that the dynamical underpinnings of human
interest are random~\cite{WH:2007,WFVM:2012,YSWAH:2012}, and this led to
the development of Markovian type of models for human interest where an
individual's history of interests (except those in the immediate past) plays
no role in his/her present action~\cite{BP:1998,CS:2007,Faginetal:2001}.
Deviations from the Markovian dynamics were
reported~\cite{MGRFM:2010,CKRS:2012,Zhaoetal:2013}. For example, a
systematic analysis of a number of big online data sets revealed that an
algebraic (power-law) scaling behavior, which is characteristic of
non-equilibrium complex systems, governs both the interevent time and event
determination statistics associated with human interest
dynamics~\cite{Barabasi:2005,Zhaoetal:2013}.
This implies that there are intrinsic dynamical rules underlying the human
interest dynamics. Three such rules were hypothesized: preferential return,
inertial effect, and exploration of new interests, and a mathematical model
was developed to explain the empirically uncovered algebraic
scaling laws~\cite{Zhaoetal:2013}.

To take into account human interest dynamics in constructing a meme
popularity model, we exploit previously studied scaling laws associated
with models~\cite{SKWB:2010,SSPTL:2012,Zhaoetal:2013}: the bursting 
characteristic of human behaviors, the algebraic distribution of the time 
required to revisit an interest, and exploration of new interest. In 
particular, we first assume that the time interval $\tau$ for an individual 
to forward the same meme to other individuals in the OSN system obeys the 
distribution $p(\tau)\sim \tau^{-\alpha}$, where $\alpha > 0$ is the algebraic 
scaling exponent that is effectively a parameter in our model. Note that 
$p(\tau)$ actually represents the bursting characteristic of human behaviors.  
Next, we assume that an individual has the probabilities $\rho$ and 
$(1-\rho)$ to forward a new and an old meme, respectively, where $\rho$ can 
be regarded as an event determination probability. The parameters $\alpha$ 
and $\rho$ can be estimated from data. 

Figure~\ref{fig:user_real} displays the user behaviors extracted from
the empirical data sets based on consideration of human interest dynamics.
The particular quantity that we examine is the activation rate or probability,
the ratio of the number of activated users at time $t$ to the number of all
users involved by this time. Figures~\ref{fig:user_real}(a-c) show, for the five
data sets, the user activation probabilities versus the normalized time $L$.
We see that, associated with the data sets {\em Douban~Book, Douban~Movie}
and {\em Douban~Music}, users exhibit a similar behavior in the activation
rates and the three curves can be well fit by a {\em sigmoid function}
[Fig~\ref{fig:user_real}(a)]. For the data set {\em Delicious} whose
activation rate curve is shown in Fig.~\ref{fig:user_real}(b), there exists
a ``step.'' We thus divide the curve into two parts and fit each with a
sigmoid function with different parameter values. For the data set {\em Weibo},
as shown in Fig~\ref{fig:user_real}(c), a single sigmoid function fits the
activation curve well. Figure~\ref{fig:user_real}(d) shows the distributions
of the individual interevent time for the five data sets, which are
approximately algebraic with the value of the exponent $\alpha$ ranging
from one to three, where the interevent time $\tau$ is defined as the time 
intervals between two consecutive actions by the same user~\cite{Barabasi:2005,
OB:2005,Dezso:2006,ZKKWH:2008,GR:2008,SKWB:2010,SSPTL:2012,Zhaoetal:2013}. 
Figures~\ref{fig:user_real}(e) and \ref{fig:user_real}(f)
display the distributions of the event determination probability $\rho$
associated with data sets {\em Delicious} and {\em Weibo}. Note that in
the data sets {\em Douban~Book, Movie} and {\em Music}, users tend to visit
or comment on an item only once, leading to a high peak in the distribution
of $\rho$ near the unity value. Table~\ref{table:fitreal} summarizes the
parameters of the sigmoid functions in Figs.~\ref{fig:user_real}(a-c) and
the parameters $\alpha$ and $\rho$ estimated directly from the data sets.
Note that, for the {\em Delicious} data set, there are two sets of values for
the parameters $B$ and $C$: one for each sigmoid function.

\begin{figure*}[htb]
\centering
\includegraphics*[width=1\linewidth]{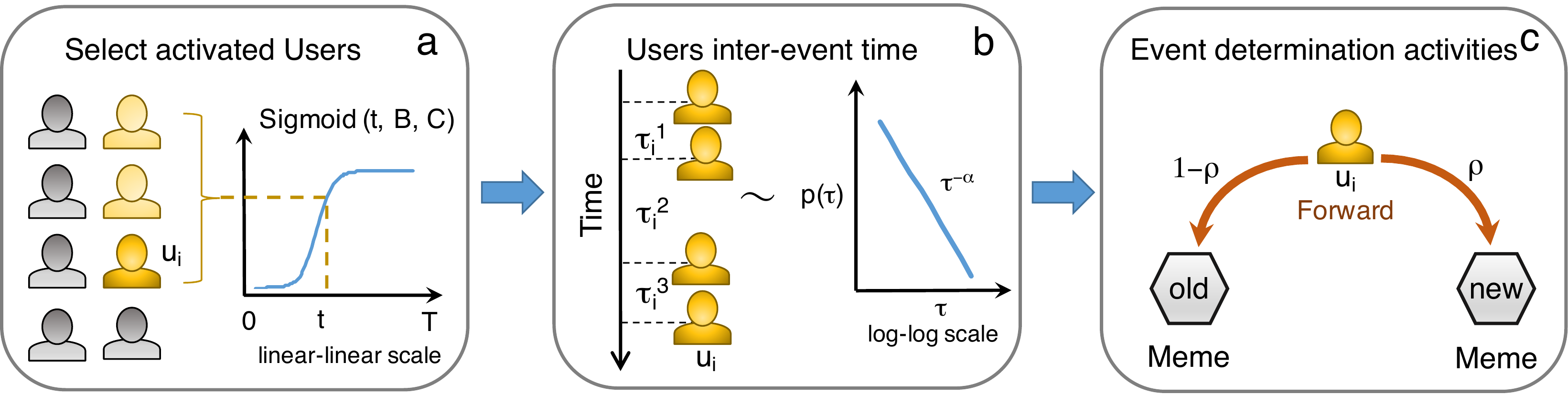}
\caption{ {\bf A schematic illustration of the proposed biomimicry based hybrid
meme diffusion model incorporating human interest dynamics and empirical
observations.} (a) At each time step $t$, a fraction of users are selected
to be activated based on a {\em sigmoid} curve. (b) For a selected user
$U_i$, the activities occur at different times according to the probability
$p(\tau)\sim \tau^{-\alpha}$, where $\tau$ is the interval between two
adjacent forwarding actions. (c) At time $t$, the selected user $U_i$ has
probability $\rho$ to forward a new meme and probability $(1-\rho)$
to forward an old meme. The model is hybrid because (1) the dynamical
evolution of the memes follows the rules of microbial cell diffusion
(biomimicry), and (2) the probabilities $p(\tau)$ and $\rho$ are from the
human interest dynamics with key parameters extracted from actual data.}
\label{fig:model}
\end{figure*}

\begin{table*}
\caption{ {\bf Estimated parameters for empirical data sets.} Parameters
associated with the sigmoid functions are $B$ and $C$ that are normalized
through $B = \tilde{B}/\mathrm{log}(T)$ and
$C = \tilde{C}/\mathrm{log}(T)$, where $\tilde{B}$ and $\tilde{C}$
are obtained from empirical data sets, $T$ is the length of the time series.
The parameters $\alpha$ and $\rho$ characterize the interevent time and
event determination activities.}
\begin{center}
\begin{tabularx}{1\textwidth}{p{0.25\textwidth}p{0.2\textwidth}p{0.2\textwidth}p{0.15\textwidth}p{0.15\textwidth}}
\hline
Data Sets & $B$ & $C$ & $\alpha$ & $\rho$ \\ \hline
Delicious & 0.24 and 0.40 & 0.80 and 0.80 & 1.73 & 0.54\\
Douban Book & 0.07 &  0.55 & 1.38  & 1.00\\
Douban Movie & 0.07 & 0.55 &  1.53 &  1.00\\
Douban Music & 0.07 & 0.55 &  1.46 & 1.00\\
Sina Weibo & 0.24 &  0.61 &  1.50 & 0.56\\
\hline
\end{tabularx}
\end{center}
\label{table:fitreal}
\end{table*}

\subsection{Construction of model for dynamical evolution of meme popularity}

Our considerations of both microbial cell evolution and human interest dynamics
in combination with the empirical observation enable a formal construction of
a model to describe the dynamical evolution of meme popularity. We start
from the basic model that includes meme diffusion and user activities.
For a fixed group of users, at each time step, a random
fraction of the users become active (enabled), and each enabled user can
post memes or forward some existing ones. In accordance with the
quantitative behaviors extracted from the empirical data sets, we assume
that the fraction of active users, $P_A(L)$, follows a sigmoid function
with parameters $B$ and $C$:
\begin{equation} \label{eq:PA}
P_A(L) \sim \frac{1}{1+ {\mathit e}^{-B\cdot(L-C)}}.
\end{equation}
For an active user, the interevent time $\tau$, the time interval
when this user decides to forward (post) a meme, follows a power law
distribution with the parameter $\alpha$. The user has 
probability $(1-\rho)$ to forward an old meme that he/she has forwarded
before and probability $\rho$ to forward a new meme. Once a meme is
created and alive on the network, its evolution is determined by the
microbial cell diffusion model. Our hybrid meme diffusion model is
illustrated in Fig.~\ref{fig:model}.

\section{Results} \label{sec:results}

\subsection{Simulation Results}

\begin{figure*}[htb]
\centering
\includegraphics*[width=1\linewidth]{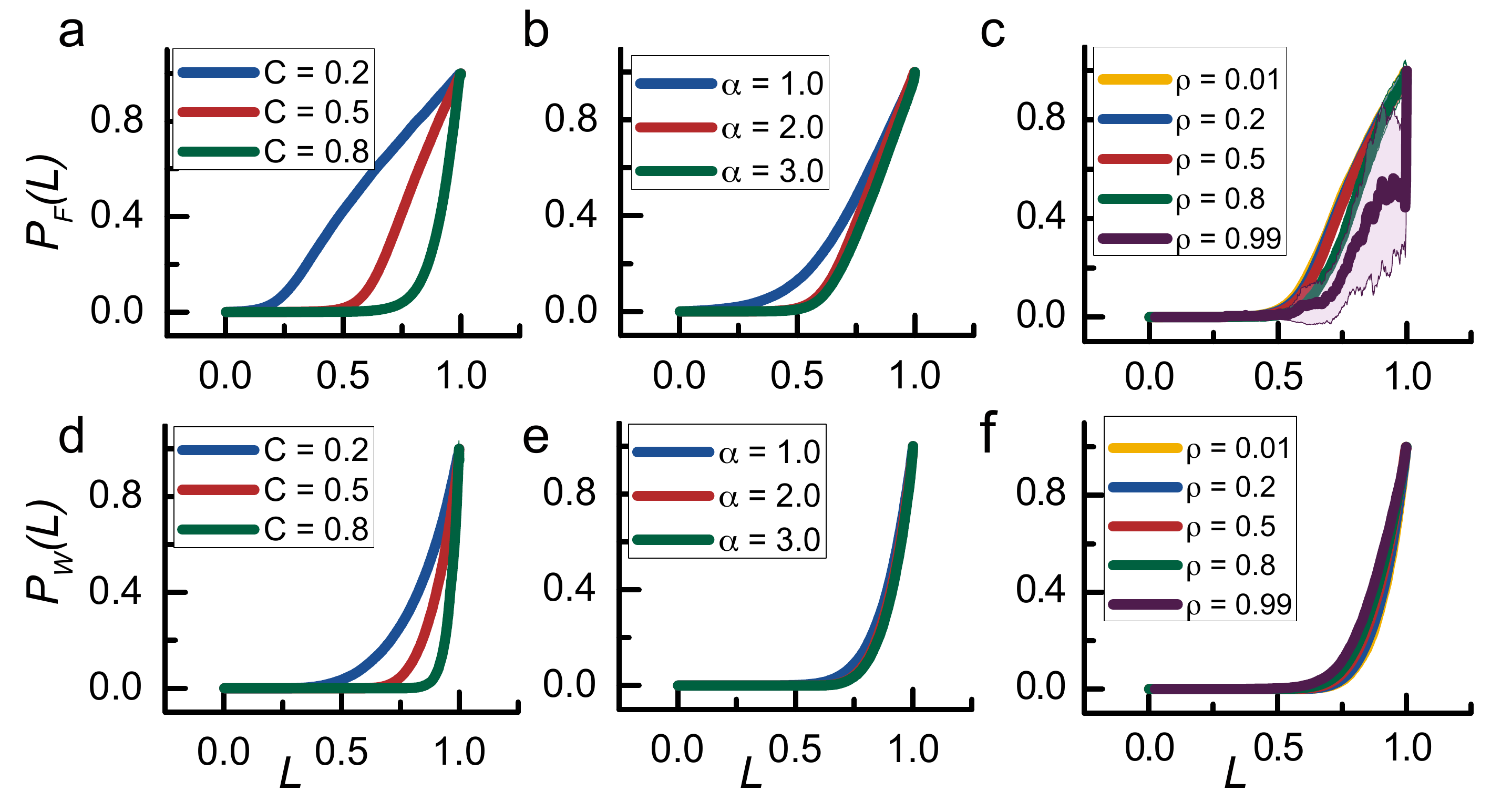}
\caption{ {\bf Time evolution of the forwarding and exclusion
probabilities generated from our hybrid model.} Normalized probabilities
for (a,d) different values of $C$ in the users activation sigmoid function,
(b,e) different values of $\alpha$ in the statistical distribution of user
interevent time, and (c,f) different values of $\rho$ in the user event
determination probabilities. In all panels, the solid curves and the associated
shadowed area indicate the average values and the error range, respectively.}
\label{fig:simulation}
\end{figure*}

To simulate the model, we consider a fixed group of $N_f =1000$ users. At
each time step $t$, $N_f\cdot P_A(t)$ randomly selected users become active.
The interevent time and event determination activities of these users
depend on two parameters: $\alpha$ and $\rho$. Specifically, for each activated
user $U_i$, we generate a time series of forwarding actions based on: (1) the
time interval $\tau$ that a forwarding action occurs follows the distribution
$p(\tau)\sim \tau^{-\alpha}$, (2) the probability for this user to forward a
new meme is $\rho$ and that to forward an old meme is $(1-\rho)$.
We then record the forwarding and exclusion probabilities, as well as the
meme popularity at each time for different choices of the parameters.
In the analysis of the empirical data, we truncate the $F$ and $W$ curves
at the peak values to exclude artificial death events caused by the finite
duration of the data sets, which typically occur at about $80\%$ of the
total duration.

Our hybrid model has four parameters: $B$ and $C$ which determine the
users' sigmoid activation rate (Eq.~\ref{eq:PA}), $\alpha$ that regulates
the interevent time intervals and $\rho$ that accounts for the event
determination activities (probability of exploring new memes). We simulate the
model for different choices of parameters to check the parameter sensitivity
of the key quantities representing the model outcome: the forwarding and
exclusion probabilities. We then estimate the four parameters directly
from the five empirical data sets.

Figure~\ref{fig:simulation} shows the simulation results of the forwarding
and exclusion probabilities for different choices of the parameters
$\alpha$, $\rho$ and $C$. For panels in a row, we modify one parameter
and compare the results, where the nominal parameter values are
$B = 0.5$, $C = 0.5$, $\alpha = 1.5$, and $\rho = 0.5$, for the reasons
that (1) a commonly used sigmoid function has $B = C = 0.5$, (2) users'
interevent time distribution exponent $\alpha$ is typically in the
range~\cite{Zhaoetal:2013} from one to three, and (3) it is reasonable to
assign equal probabilities for both old and new memes when the event
determination activities are not known, leading to the choice $\rho = 0.5$.
A key result of Fig.~\ref{fig:simulation} is that the forwarding
and exclusion probabilities are sensitive to user actions. For the sigmoid
activation function, a small value of $C$ means that the users get activated
at an early time, resulting in increased forwarding and exclusion
probabilities at the early time, as shown in Figs.~\ref{fig:simulation}(a)
and \ref{fig:simulation}(d). In terms of the interevent time distribution,
a large value of $\alpha$ means that the distribution is more concentrated,
indicating that many users have short interevent time. This
leads to a slow growth in the forwarding probability at an early time, as
shown in Fig.~\ref{fig:simulation}(b). While the increment is not dramatic
as compared with that in the forwarding probability, there is a slight delay
in the rising of the exclusion probability in early time when the value of
$\alpha$ is large, as shown in Fig.~\ref{fig:simulation}(e). We note that the
forwarding probability changes slightly with different values of the event
determination parameter $\rho$. For example, a large value of $\rho$, which
leads to a higher probability to forward new memes, results in a delayed
increment in the forwarding probabilities, and the high probability of new
memes indicates a small number of forwarding events per meme on the average,
which causes the early increment of the exclusion probabilities. In all six
panels, the amplitudes of increase in the forwarding probability with different
parameters values are always larger than those of the exclusion probabilities.
This is because in simulations, a meme is regarded as being excluded only
after the last recorded forwarding actions so that the influence of user
actions on the overwriting action is postponed.

\begin{figure*}[htb]
\centering
\includegraphics*[width=1\linewidth]{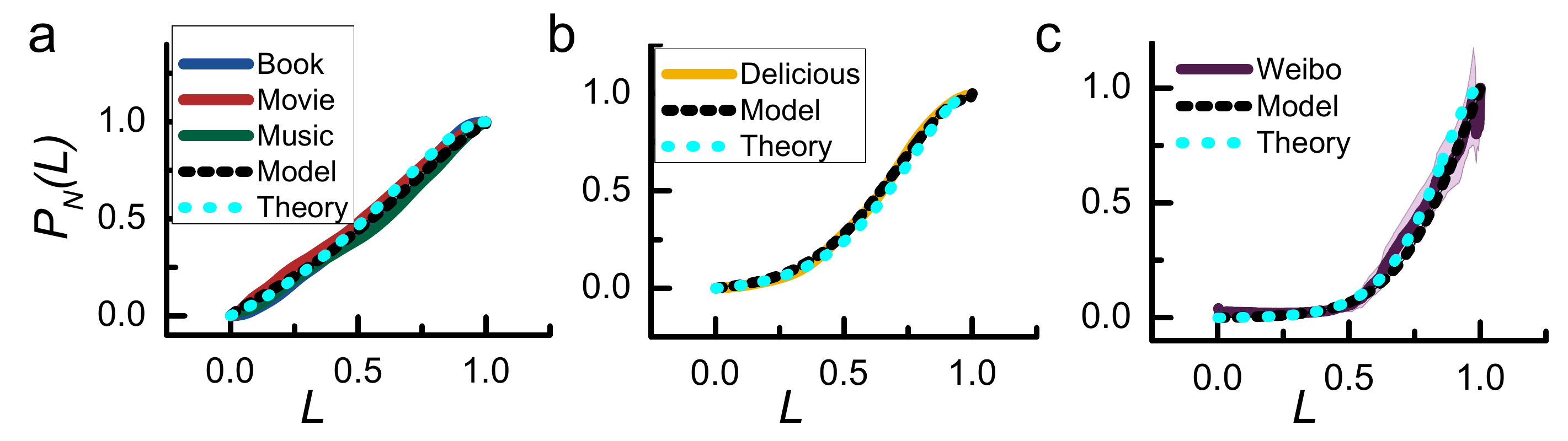}
\caption{ \textbf{Test of the predictive power of the proposed model:
comparison between model predicted and real grow curves of meme popularity.}
Dashed and solid curves represent model prediction and the results from the
real data, respectively: (a) linear growth of meme popularity for the
{\em Douban~Book, Douban~Movie} and {\em Douban~Music} OSN systems,
(b) an ``S-shape'' type of growth behavior from the data set {\em Delicious},
and (c) an approximately exponential growth curve for the data set
{\em Weibo}. The four key model parameters $B$, $C$, $\alpha$, and $\rho$
are estimated from data. For the data set {\em Delicious}, the empirical
fitting requires two sigmoid functions for the activation rate with
different values of parameters $B$ and $C$ (see Tab.~\ref{table:fitreal}).
However, our hybrid model requires a single sigmoid function only
(for $B = 0.15$ and $C = 0.5$) to generate the meme popularity growth
curve that agrees well with the empirical results. The remarkable feature
is that, regardless of the characteristically different growth behaviors
in the meme popularity associated with diverse OSN systems, both model
simulations and analytic solutions are capable of accurate prediction.
Parameters used for solving Eq.~(\ref{eq:final}) are $T=1000$ and $dt=0.01$.
The model, analytic and real results have passed the standard 
Kolmogorov-Smirnov (KS) test~\cite{CF:2009} with $D = 0.05$.}
\label{fig:realNsim}
\end{figure*}

To validate our model, we set its four essential parameters to the values
estimated from real data and compare the model predicted dynamical evolution
of meme popularity with the real one. We do this for all five big OSN data sets,
as shown in Fig.~\ref{fig:realNsim}, where the growth curves of the normalized
meme popularity with time [${{\rm{P}}_N}(t) \to {{\rm{P}}_N}(L)$] from model
and data are displayed. 
The remarkable feature in Fig.~\ref{fig:realNsim} is that, regardless of the 
disparity in the nature of the data sets and regardless of the characteristic 
differences in the growth dynamics of meme popularity, our model predicts 
behaviors that agree with the actual behavior accurately at a detailed level. 
In particular, for the data sets {\em Douban~Book, Movie} and {\em Music} 
in Fig.~\ref{fig:realNsim}(a), the meme popularity grows linearly with time 
and our model predicts this behavior precisely. 
For the data set {\em Delicious}, the meme popularity growth curve exhibits 
an ``S-shape'' feature, which is characteristically different from the linear 
behavior in Fig.~\ref{fig:realNsim}(a), but the model prediction based on a 
single set of parameter values $(B,C,\alpha,\rho)$, which is in good agreement 
with the empirical result, captures this distinct feature unequivocally. For 
the growth behavior in Fig.~\ref{fig:realNsim}(c) from the data set 
{\em Weibo}, the meme popularity exhibits an exponential behavior over time, 
which is predicted by our model.

We test the results from our mathematical analysis of the hybrid meme 
polarity growth model in Sec.~\ref{subsec:theory}. 
Numerical solutions of Eq.~(\ref{eq:final}) are shown by
the cyan dashed lines in Fig.~\ref{fig:realNsim}. For both analytic and
numerical solutions, the parameters ($B$, $C$) associated with the users
activation function and the event determination parameter $\rho$ are key to
determining the shape of the meme popularity growth curves.
We find that fluctuations of the parameters about their ``exact'' values
can be tolerated in our model without affecting its prediction. That is, our
hybrid model not only predicts successfully the meme popularity growth
dynamics, but also is robust against parameter inaccuracy and uncertainties.


\subsection{Mathematical analysis of meme popularity growth model} 
\label{subsec:theory}

From Eq.~(\ref{eq:Nt}) and Fig.~\ref{fig:illustration}, we see that
the meme population at time $t$ depends on the numbers of the survived
and forwarded memes, which can be described by the following relation:
\begin{equation*}
N(t)=S(t-1)+F(t-1)-W(t)+F_A(t),
\end{equation*}
where $F_A(t)$ is the number of newly added memes at time $t$. The
recursive relation can be approximated by the following differential
equation:
\begin{equation} \label{eq:ODE_N}
\frac{dN(t)}{dt}=F_A(t)-W(t).
\end{equation}
In simulations, at each time step, we add some users into the system according
to the sigmoid function. Based on natural human behavior as illustrated in
Fig.~\ref{fig:model}, a subset of the newly added users contribute to new
memes at the same time. The fraction of the added new memes is $\rho$, so
the number of newly added memes is
\begin{equation} \label{eq:activation}
F_A(t)=\rho N_i \cdot \frac{1}{1+\mathrm {exp}
\big({-\tilde B(t-\tilde C)}\big)}.
\end{equation}
After the last forwarding event, a meme is
regarded as being overwritten or excluded. Since the user interevent time
intervals follow the distribution $p(\tau) \sim \tau ^{-\alpha}$, it is
reasonable to assume that all users' last interevent interval follows
the same distribution, so does the memes' last interevent interval.
The death term $W(t)$ in Eq.~(\ref{eq:ODE_N}) depends on the total number
of memes added into the system, which can be written as:
\begin{equation} \label{eq:W}
W(t)=\int_0^t \frac{\rho N_i}
{1+\mathrm {exp}\big(-\tilde B(t'-\tilde C)\big)}dt'\cdot (T-t)^{-\alpha}.
\end{equation}
where $T$ is the final observation time, as shown in Fig.~\ref{fig:model}(a).
Substituting Eqs.~(\ref{eq:activation}) and~(\ref{eq:W}) to
Eq.~(\ref{eq:ODE_N}), we obtain
\begin{eqnarray} \label{eq:CF_ODE}
	\frac{dN(t)}{dt} & = & \rho N_i \cdot\frac{1}{1+\mathrm{exp}
\big(-\tilde B(\hat{t} - \tilde C)\big)} \\ \nonumber
	& - & \int_0^{\hat{t}}
\frac{\rho N_i}{1+\mathrm {exp}\big(-\tilde B(t'-\tilde C)\big)}dt'
\cdot(\hat{T}-\hat{t})^{-\alpha},
\end{eqnarray}
where $\hat{t}$ and $\hat{T}$ are normalized time defined as
\begin{eqnarray}
\nonumber
\hat{t} & = & t\cdot\mathrm{log}(T)/T, \\ \nonumber
\hat{T} & = & \mathrm{log}(T).
\end{eqnarray}
The integral in Eq.~(\ref{eq:CF_ODE}) can be calculated analytically,
leading to
\begin{widetext}
\begin{equation} \label{eq:final}
\begin{split}
\frac{dN(t)}{dt}&=\rho N_i\cdot\frac{1}{1+\mathrm{exp}
\big(-\tilde B(\hat{t}- \tilde C)\big)}-\frac{\rho N_i}
{\tilde B}\cdot\bigg(\tilde B\cdot\hat{t}-\mathrm{log}
\big(1+\mathrm{exp}(\tilde B\cdot \tilde C)\big)
+\mathrm{log}\Big(1+\mathrm{exp}\big(\tilde B
\cdot(\tilde C-t)\big)\Big)\bigg)\cdot(\hat{T}-\hat{t})^{-\alpha}.
\end{split}
\end{equation}
\end{widetext}
The derivation of the analytic prediction of meme popularity growth,
Eq.~(\ref{eq:final}), relies on
approximations such as taking the average of the growth and exclusion
processes. For typical OSN systems where the meme and user numbers are
large, the approximations are reasonable. Solutions of Eq.~(\ref{eq:final})
for different choices of the parameters $B$, $C$, $\alpha$ and $\rho$
adopted from Table~\ref{table:fitreal} are typically curves with both
increasing and decreasing phases. To be consistent with the simulation
settings, we truncate curves of the analytic solutions at the point
where they start to decrease. As shown in Fig.~\ref{fig:realNsim},
there is a good agreement among the three types results: analytic prediction,
results from direct simulation of the hybrid model, and those from the
empirical data.

A remarkable feature of Fig.~\ref{fig:realNsim} is that, associated with
the three types of results there are characteristically different growth
behaviors: linear, ``S-shape'' and exponential, depending on the choices
of the parameters $B$, $C$, and $\rho$. To gain a qualitative understanding,
we examine the effects of these parameters on the meme popularity dynamics.
In particular, $B$ and $C$ are parameters in the activation sigmoid function,
where $B$ controls the shape of the function and $C$ modulates the value of
$L$ when the activation rate reaches the value of $0.5$ in Eq.~(\ref{eq:PA}).
A large $C$ value will then result in a late arrival of the half of the
total meme popularity, and a large $B$ value will make the slope of the
sigmoid function steeper. A large increasing rate in the popularity can
be expected in the region where the normalized time $L$ is close to the
$C$ value. As a result, large values of $B$ and $C$ will delay the onset of
the popularity growth to a later time, but when growth does start, it does so
at a large rate. This explains the exponentially growing behavior in
Fig.~\ref{fig:realNsim}(c). For moderate values of $B$ and $C$ (e.g.,
$B$ around $0.15$ and $C$ about $0.5$), the onset of popularity growth
occurs at an earlier time, reaches the maximum rate at a later time when
the decrease due to exclusion and the increase due to the newly added
memes are balanced, leading to an ``S-shape'' type of growth curves, as
illustrated in Fig.~\ref{fig:realNsim}(b). Finally, for small $B$ and
moderate $C$ values, the slope of the sigmoid function is small, leading
to an approximately linear growth behavior in the meme popularity.

Note that, the second term in Eq.~(\ref{eq:final}) represents the number
of excluded memes. For small $\hat{t}$, this exclusion term adds only a
small modification to meme's popularity. For $\hat{t} \sim \hat{T}$, this
term plays an important role in the growth behavior. The parameter $\rho$
does not affect the shape of the growth curve and, in fact, its effect is
quite straightforward: a large (small) value of $\rho$ leads to a large
(small) absolute popularity value. In a real system, for small values of
$\rho$ (e.g., $0.02$), stochastic fluctuations will play a non-negligible
role in the growth behavior, which can be seen through a comparison of
Figs.~\ref{fig:realNsim}(a-c).

\section{Discussion} \label{sec:discussion}

Recent years have witnessed an unprecedentedly rapid growth of OSN systems.
These systems have become ubiquitous in the modern society with the tendency
to eventually replace many of the traditional social networks. To understand
the dynamics of and the dynamical processes on OSN systems is of 
importance to the well being of the human society. There have
been efforts in analyzing OSN systems in the past few years~\cite{OR:2010,
TWL:2012,Zhaoetal:2013,WHRWL:2014,LKML:2014,KPH:2014,KB:2014,
GWOL:2014,Ribeiro:2014,SL:2015,ZYMH:2016,GOBM:2016}. A common approach is
to search for certain statistical or scaling relations through (big) data
analysis, and then to articulate a quantitative model to reproduce the
specific scaling laws (often power-law scaling). While this approach has
indeed yielded great insights into the dynamics of specific OSN systems,
the inconvenient truth is that these systems are diverse and exhibit
characteristically different behaviors even for a single quantity of
interest.

A ubiquitous phenomenon with general interest is the growth dynamics
of meme popularity in OSN systems. Our analysis of empirical data from
diverse OSN systems reveals a lack of common growth behavior:
depending on the system the growth dynamics can be linear, ``S-shaped,''
or exponential. We ask the challenging question: can a single,
universal model be articulated to capture and predict the characteristically
different growth behaviors of meme popularity? Naturally, in order to produce
distinct dynamics, the model should contain free parameters whose
values depend on the specific system and should be estimated from data. In
spite of this, previous phenomenological models were not generally applicable
to diverse OSN systems. We are thus motivated to articulate our model based
on the fundamental dynamical elements underlying the growth and spread
of meme popularity. We find biomimicry to be highly inspirational, effective,
and useful for achieving this goal. Specifically, we realize that the
dynamics of cell evolution in microbial ecology bear similarities to the
dynamics of meme popularity.

In microbial ecology, at any time a cell can experience one of the
three events: division, death, or survival. Likewise, in OSN systems,
at any time a meme can be forwarded, overwritten, or can survive.
In both contexts, each event is associated with a probability that is a
function of time with free parameters. The close resemblance between
cell and meme popularity growth dynamics provides the microscopic base
for a population model of memes. Through data analysis, we find that
the key probability functions have the shape of a sigmoid function, with
two free parameters that can be determined from data. However, relying
solely on the bio-inspired growth model is not sufficient to capture the
detailed dynamical behaviors of meme popularity growth, as the OSN systems
involve human behaviors and can thus be significantly more complex than just
biological cell growth. For meme popularity, it is natural and reasonable
to attribute the modeling element beyond cell growth to human interest
dynamics, for which models have been developed recently. Incorporating the
ingredient of human interest into basic cell growth, we end up with a hybrid
model that contains four free parameters - all determinable from data. With
the parameter values so estimated, we demonstrate that our model has the
ability to predict the detailed, characteristically distinct growth
dynamics of meme popularity in diverse OSN systems. To our knowledge, this
is the first time that a universal model has been successfully developed to
capture and predict the dynamical evolution of a broadly interested entity
characteristic of modern social networking systems.

We remark that, in spite of the large meme varieties, it is still possible to
construct a general model for the meme growth dynamics through the actions of
forwarding (posting), staying survival, and being overwritten (exclusion).
To capture the essential dynamics while making the model simple and analyzable, 
additional factors such as friendship reciprocity, visibility, intrinsic 
interaction among users, or other regular operations on past memes were not 
taken into account in our model. More specifically, in the real world, a 
user's activity is the overall result of its own intrinsic motivation and 
the influence of its peers~\cite{PBHMGB:2018}. Our current model does not 
have the ability to deal with this complicated issue mainly because not a 
single data set includes detailed information about these additional factors 
(although some data sets contain limited information about user behaviors).
Moreover, one focused aspect of our model is the effects of user intrinsic 
activities on the meme popularity. To make this possible, we have neglected 
user-user interactions or the influences of peers, which are certainly 
important for further development of the model.

From the control point of view, our hybrid model leads to a new framework
to understand the general mechanisms underlying the dynamics of OSN systems
and how they may be manipulated or harnessed. Regardless of the specific
features in the OSNs studied in this paper [from bookmarking shared
networks ({\em Delicious}) to interest discovered media ({\em Douban})
and microblogging system ({\em Weibo})], they all share the same
``microscopic'' dynamical elements (e.g., forwarding and exclusion) that
results in certain rapid growth behavior. 
It is conceivable that the growth dynamics can be controlled through
perturbations to these dynamical elements. This may lead to network design
that, for instance, optimizes meme populations. It is also possible to
optimize the robustness or resilience of information spreading in OSN
systems to external attacks through modulating the forwarding and
exclusion probabilities. Taken together, our work uncovers the fundamental
principles underlying meme popularity in OSNs through the approach of
biomimicry in combination with insights from human behavior dynamics,
which may shed lights on further development of social networks and
outstanding issues such as classification, robustness, optimization,
and control.

\section*{Acknowledgement}

The first two authors contributed equally to this work.
We thank Dr.~J.-M. Huang for providing the Douban data set and Dr.~X.-F. Wang
for providing the Weibo data set. We would like to acknowledge support from
the Vannevar Bush Faculty Fellowship program sponsored by the Basic Research
Office of the Assistant Secretary of Defense for Research and Engineering and
funded by the Office of Naval Research through Grant No.~N00014-16-1-2828.
ZDZ was partially supported by the NSF of China under Grant No.~61603074.

\section*{appendix}
\setcounter{figure}{0}
\renewcommand\thefigure{A\arabic{figure}}

\subsection{Results from empirical data sets}

\begin{figure*}[h!]
\centering
\includegraphics[width=\linewidth]{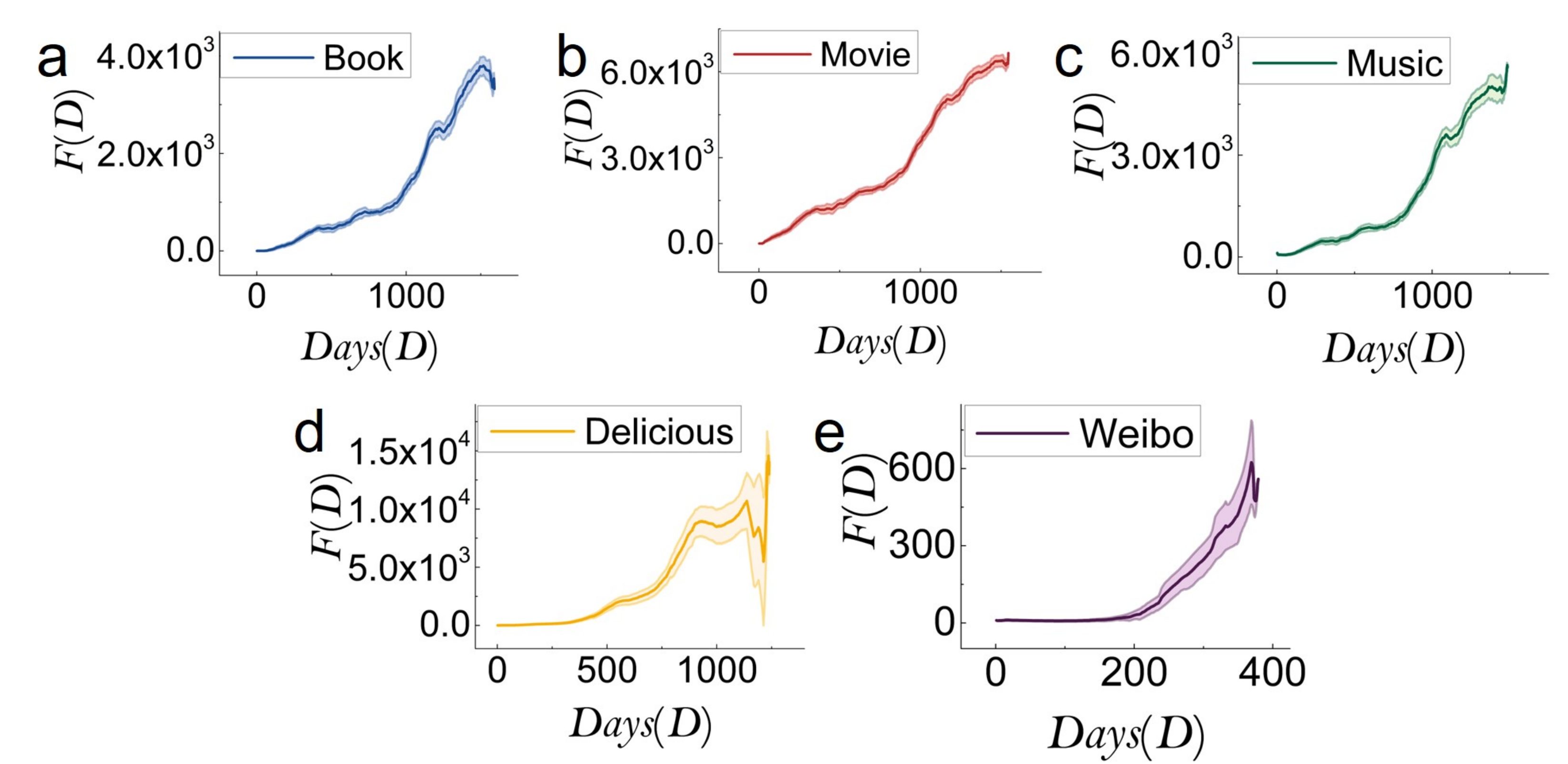}
\caption{{\bf Number of forwards from the five empirical data sets studied.} 
In each panel, the line and the shadow area represent the average number of 
forwards and the error, respectively.}
\label{fig:realF}
\end{figure*}

\begin{figure*}[h!]
\centering
\includegraphics[width=\linewidth]{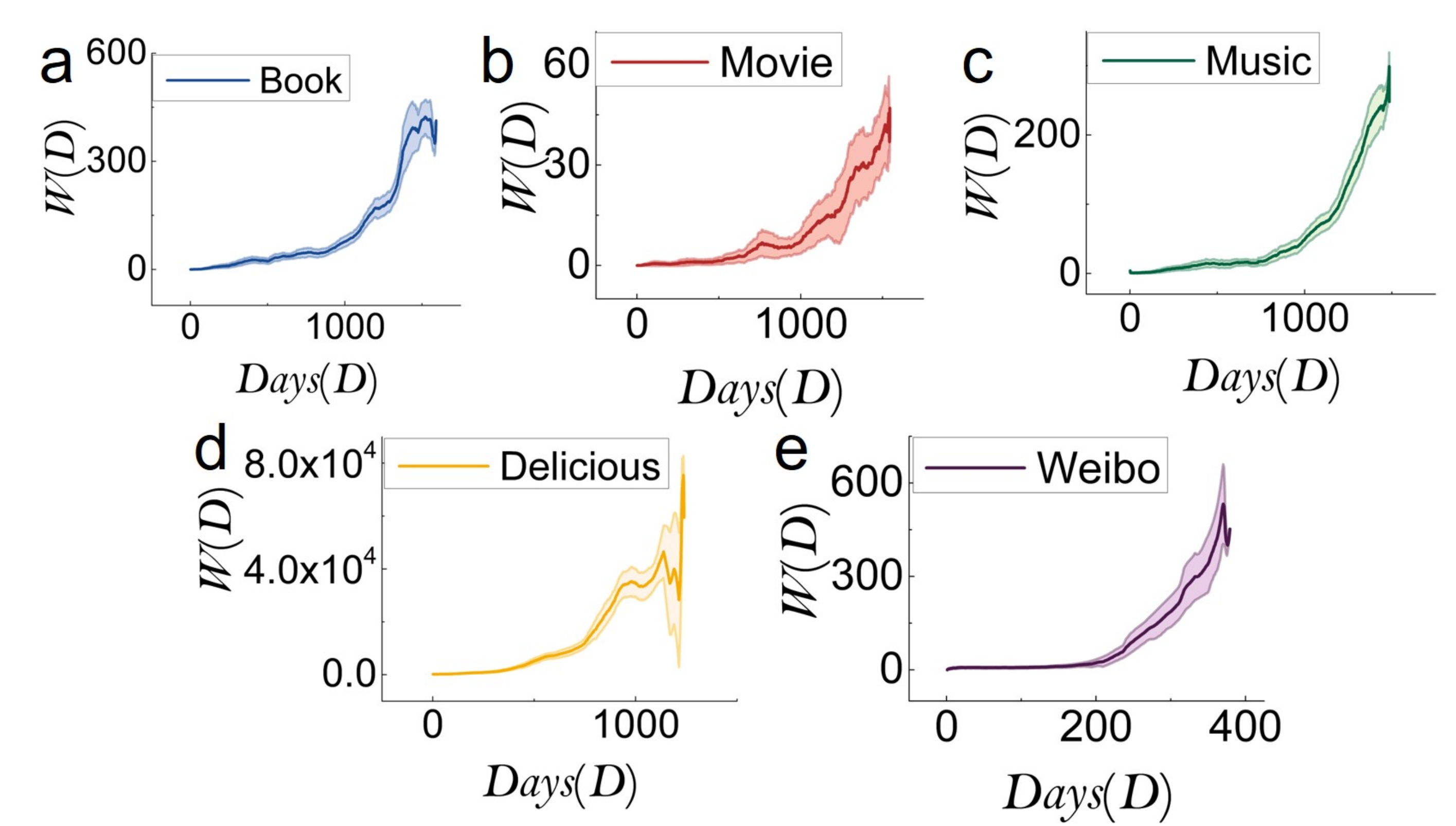}
\caption{{\bf Number of overwrites in the five empirical data sets studied}. 
In each panel, the line and the shadow area represent the average number of 
forwards and the error, respectively.} 
\label{fig:realW}
\end{figure*}

To gain quantitative insights into the number of occurrence for each event 
in the empirical data sets, we illustrate the results on the numbers of 
forwards and overwrites in Figs.~\ref{fig:realF} and 
\ref{fig:realW}, respectively.

\subsection{Effects of parameter variations}

\begin{figure*}[h!]
\centering
\includegraphics[width=\linewidth]{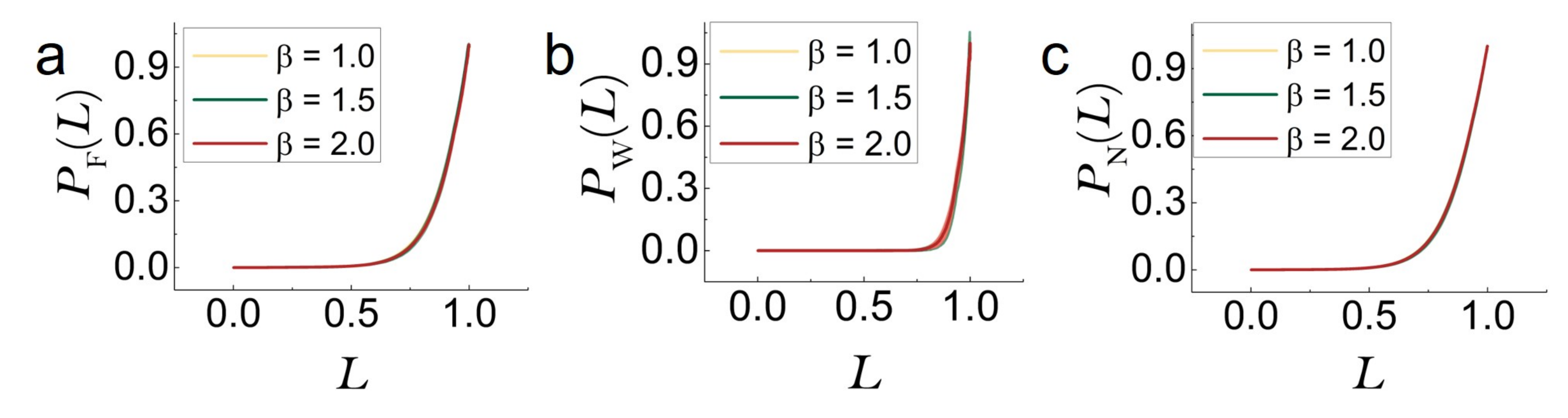}
\caption{{\bf Effects of the value of $\beta$ on the dynamical rates}. 
(a-c) The forwarding, overwritten and popularity rates for different 
values of parameter $\beta$ associated with modeling of users' intra-event 
process. Small variations of $\beta$ have little effect on the rates.}
\label{fig:betaSim}
\end{figure*}

As discussed in the main text, users' intra-event activities can be 
described by two probabilities: $\rho\cdot n^{-\beta}$ to 
forward an old meme and $(1-\rho\cdot n^{-\beta})$ to 
forward a new meme, where $n$ is the number of hopping events and $\beta$ 
is an index (see user interest modeling in the main text). We find that,
for different values of $\beta$, the forwarding, overwritten and popularity 
rates hardly change, as shown in Fig~\ref{fig:betaSim}, 
justifying the use of a single parameter $\rho$ to characterize users' 
intra-event activities.

\begin{figure*}[h!]
\centering
\includegraphics[width=\linewidth]{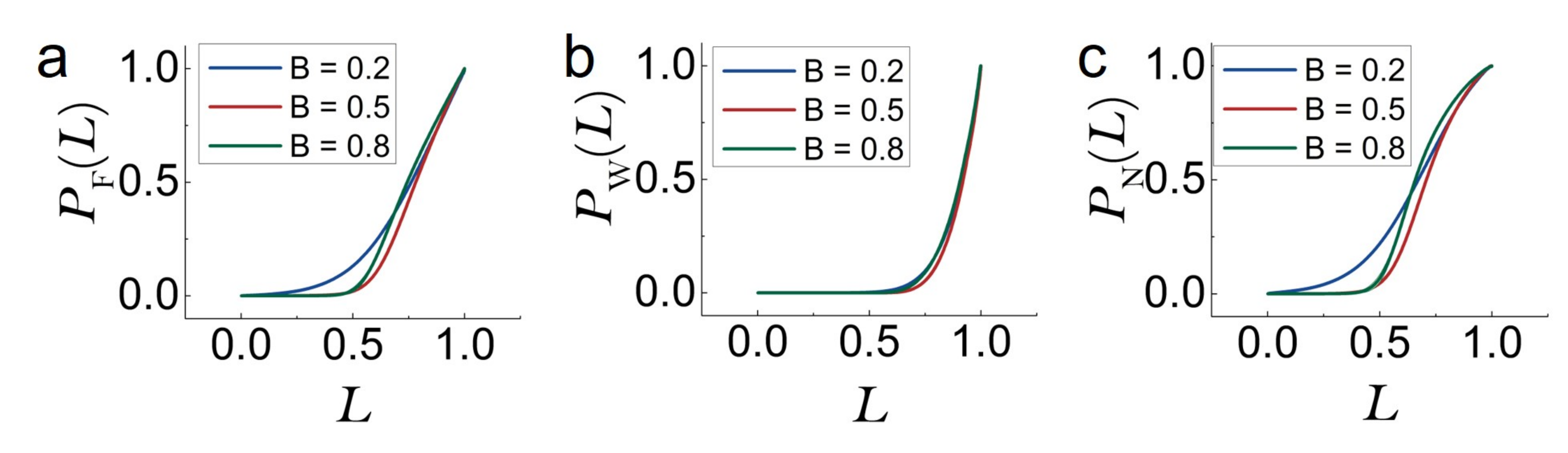}
\caption{{\bf Effects of the value of $B$ on the dynamical rates}.
(a-c) The forwarding, overwritten and popularity rates for different 
values of parameter $B$ associated with modeling of users' intra-event 
process. Small variations of $B$ have little effect on the rates.}
\label{fig:BSim}
\end{figure*}

We also examine the effects of small variations in the parameter $B$ 
in the users activation Sigmoid function [Eq.~(2) in main text] 
on the forwarding, overwritten and popularity rates, as shown in  
Fig.~\ref{fig:BSim}. While varying the value of $B$ can 
affect the rates of different events (especially the forwarding rate), 
the effects negligible as compared with, e.g., those of varying the 
parameter $C$ as shown in panels (a) and (d) in Fig.~5 of the main text. 
We thus fix $B=0.5$ in our model analysis and simulations.

\subsection{Regularities in the empirical data sets}

\begin{figure*}[h!]
\centering
\includegraphics[width=\linewidth]{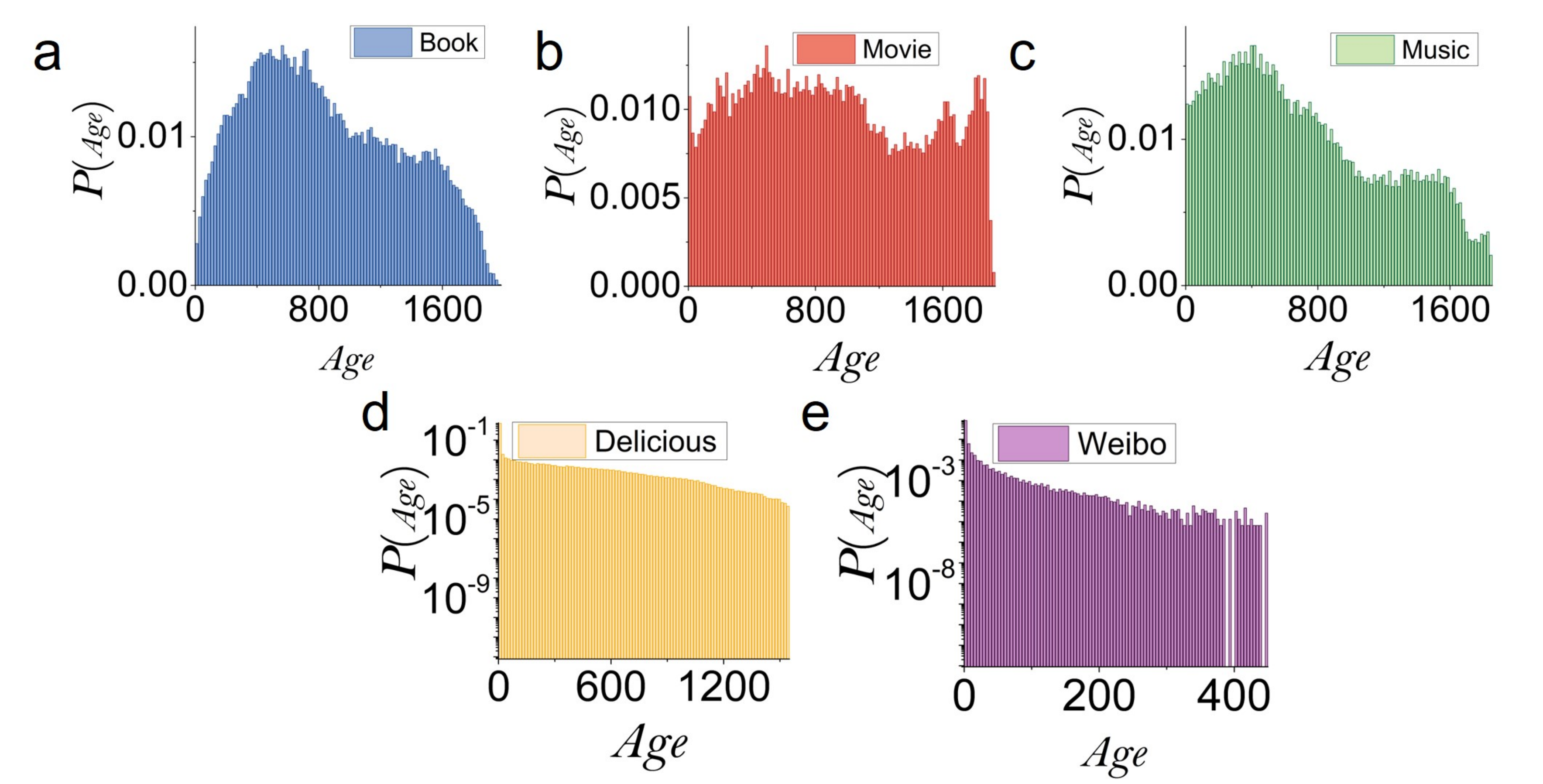}
\caption{{\bf Meme age distribution in the five empirical data sets studied.}
(a-c) Distributions of meme's age (in days) in {\em Douban Book, Movie} and 
{\em Music} data sets, respectively, on a linear-linear scale. 
(d,e) Meme age distribution in {\em Delicious} and {\em Weibo} on a 
logarithmic-logarithmic scale.}
\label{fig:ageDistr}
\end{figure*}

In population modeling of both cell growth in microbial ecology and 
meme popularity in  social networking systems, a factor of consideration 
is cell or meme age~\cite{SAKWS:2013,ZFB:2014,GWOL:2014,GOBM:2016,CG:2016}. 
In meme modeling, ``age'' is defined as the time duration between a meme's 
first appearance and death. Figure~\ref{fig:ageDistr} shows  
the distribution of meme's age for all the empirical data sets studied.
We see that, for data sets {\em Douban Book, Movie} and {\em Music}, the
age distributions are relatively homogeneous, where most memes have the 
age of $400$ to $1200$ days. For {\em Delicious} and {\em Weibo}, memes 
have a heterogeneous age distribution, where many memes have short duration. 
The difference in the age distribution suggests that the corresponding OSN 
systems do possess different structures, but our framework is flexible and 
general enough to make unified modeling of popularity diffusion dynamics of 
different OSNs possible. 

\begin{figure*}[h!]
\centering
\includegraphics[width=0.6\linewidth]{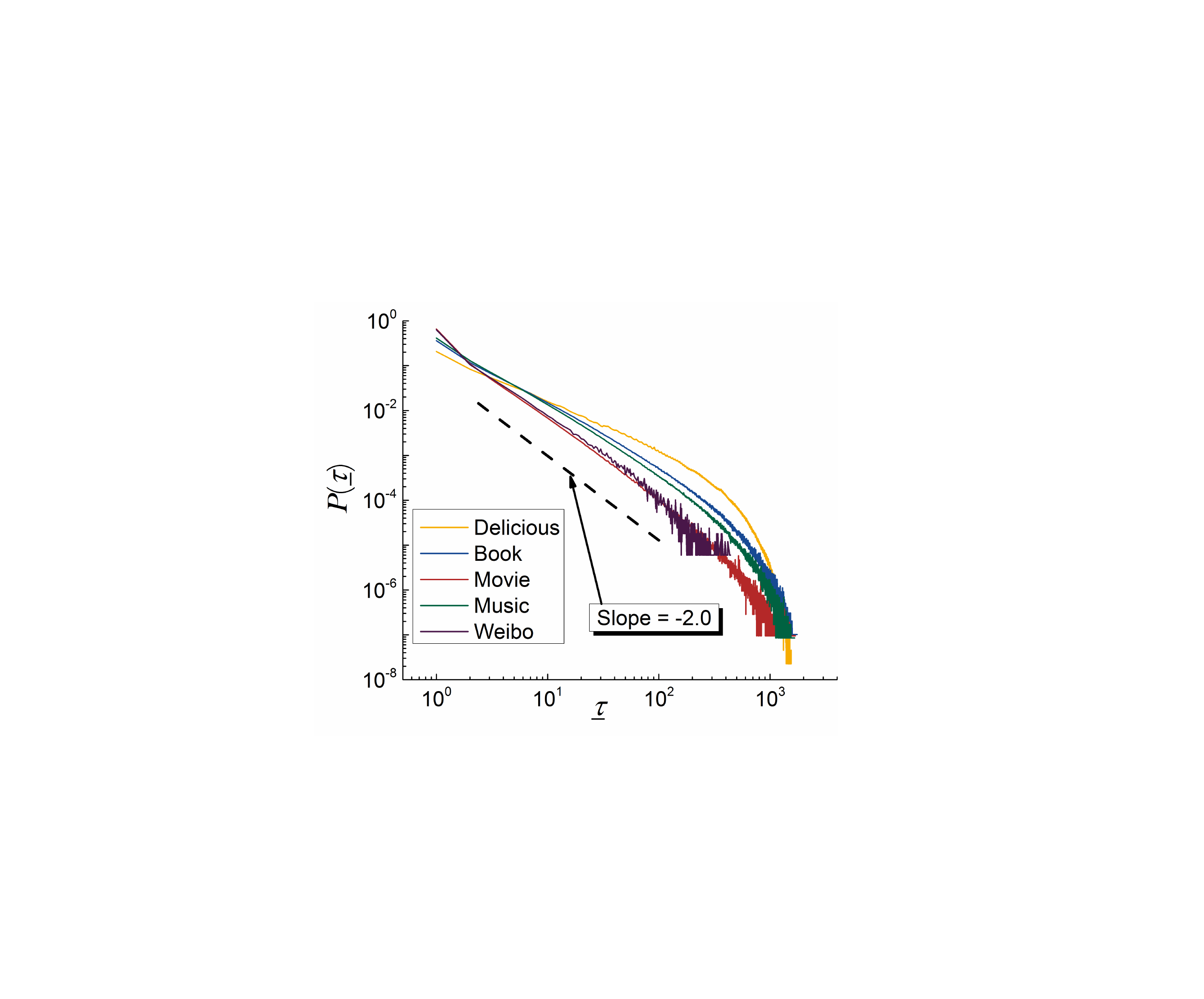}
\caption{{\bf Distributions of meme's forwarding intervals for the five 
empirical data sets studied.} Yellow, blue, red, green and purple solid 
curves represent the power-law distributions of meme forwarding intervals 
for {\em Delicious}, {\em Douban~Book}, {\em Douban~Movie}, {\em Douban~Music} 
and {\em Weibo}, respectively: 
$p(\underline{\tau}) \sim \underline{\tau}^{-\gamma}$.}
\label{fig:memeTau}
\end{figure*}

Heterogeneity in the OSN systems is also reflected in the distribution 
of the time interval $\underline{\tau}$ for memes, as shown in  
Fig.~\ref{fig:memeTau} for the five data sets studied.
We see that most memes have short time intervals (gaps) and only a few 
memes have large intervals.

\begin{thebibliography}{69}%
\makeatletter
\providecommand \@ifxundefined [1]{%
 \@ifx{#1\undefined}
}%
\providecommand \@ifnum [1]{%
 \ifnum #1\expandafter \@firstoftwo
 \else \expandafter \@secondoftwo
 \fi
}%
\providecommand \@ifx [1]{%
 \ifx #1\expandafter \@firstoftwo
 \else \expandafter \@secondoftwo
 \fi
}%
\providecommand \natexlab [1]{#1}%
\providecommand \enquote  [1]{``#1''}%
\providecommand \bibnamefont  [1]{#1}%
\providecommand \bibfnamefont [1]{#1}%
\providecommand \citenamefont [1]{#1}%
\providecommand \href@noop [0]{\@secondoftwo}%
\providecommand \href [0]{\begingroup \@sanitize@url \@href}%
\providecommand \@href[1]{\@@startlink{#1}\@@href}%
\providecommand \@@href[1]{\endgroup#1\@@endlink}%
\providecommand \@sanitize@url [0]{\catcode `\\12\catcode `\$12\catcode
  `\&12\catcode `\#12\catcode `\^12\catcode `\_12\catcode `\%12\relax}%
\providecommand \@@startlink[1]{}%
\providecommand \@@endlink[0]{}%
\providecommand \url  [0]{\begingroup\@sanitize@url \@url }%
\providecommand \@url [1]{\endgroup\@href {#1}{\urlprefix }}%
\providecommand \urlprefix  [0]{URL }%
\providecommand \Eprint [0]{\href }%
\providecommand \doibase [0]{http://dx.doi.org/}%
\providecommand \selectlanguage [0]{\@gobble}%
\providecommand \bibinfo  [0]{\@secondoftwo}%
\providecommand \bibfield  [0]{\@secondoftwo}%
\providecommand \translation [1]{[#1]}%
\providecommand \BibitemOpen [0]{}%
\providecommand \bibitemStop [0]{}%
\providecommand \bibitemNoStop [0]{.\EOS\space}%
\providecommand \EOS [0]{\spacefactor3000\relax}%
\providecommand \BibitemShut  [1]{\csname bibitem#1\endcsname}%
\let\auto@bib@innerbib\@empty
\bibitem [{\citenamefont {Holme}\ and\ \citenamefont {Newman}(2006)}]{HN:2006}%
  \BibitemOpen
  \bibfield  {author} {\bibinfo {author} {\bibfnamefont {P.}~\bibnamefont
  {Holme}}\ and\ \bibinfo {author} {\bibfnamefont {M.~E.~J.}\ \bibnamefont
  {Newman}},\ }\bibfield  {title} {\enquote {\bibinfo {title} {Nonequilibrium
  phase transition in the coevolution of networks and opinions},}\ }\href@noop
  {} {\bibfield  {journal} {\bibinfo  {journal} {Phys. Rev. E}\ }\textbf
  {\bibinfo {volume} {74}},\ \bibinfo {pages} {056108} (\bibinfo {year}
  {2006})}\BibitemShut {NoStop}%
\bibitem [{\citenamefont {Stehl\'e}, \citenamefont {Barrat},\ and\
  \citenamefont {Bianconi}(2010)}]{SBB:2010}%
  \BibitemOpen
  \bibfield  {author} {\bibinfo {author} {\bibfnamefont {J.}~\bibnamefont
  {Stehl\'e}}, \bibinfo {author} {\bibfnamefont {A.}~\bibnamefont {Barrat}}, \
  and\ \bibinfo {author} {\bibfnamefont {G.}~\bibnamefont {Bianconi}},\
  }\bibfield  {title} {\enquote {\bibinfo {title} {Dynamical and bursty
  interactions in social networks},}\ }\href@noop {} {\bibfield  {journal}
  {\bibinfo  {journal} {Phys. Rev. E}\ }\textbf {\bibinfo {volume} {81}},\
  \bibinfo {pages} {035101} (\bibinfo {year} {2010})}\BibitemShut {NoStop}%
\bibitem [{\citenamefont {Cha}\ \emph {et~al.}(2010)\citenamefont {Cha},
  \citenamefont {Haddadi}, \citenamefont {Benevenuto}, \citenamefont
  {Gummadi},\ and\ \citenamefont {Krishna}}]{CHBG:2010}%
  \BibitemOpen
  \bibfield  {author} {\bibinfo {author} {\bibfnamefont {M.-Y.}\ \bibnamefont
  {Cha}}, \bibinfo {author} {\bibfnamefont {H.}~\bibnamefont {Haddadi}},
  \bibinfo {author} {\bibfnamefont {F.}~\bibnamefont {Benevenuto}}, \bibinfo
  {author} {\bibfnamefont {K.}~\bibnamefont {Gummadi}}, \ and\ \bibinfo
  {author} {\bibfnamefont {P.}~\bibnamefont {Krishna}},\ }\bibfield  {title}
  {\enquote {\bibinfo {title} {Measuring user influence in twitter: The million
  follower fallacy},}\ }in\ \href@noop {} {\emph {\bibinfo {booktitle} {4th
  International AAAI Conference on Weblogs and Social Media (ICWSM)}}},\
  Vol.~\bibinfo {volume} {14}\ (\bibinfo {year} {2010})\ p.~\bibinfo {pages}
  {8}\BibitemShut {NoStop}%
\bibitem [{\citenamefont {Onnela}\ and\ \citenamefont
  {Reed-Tsochas}(2010)}]{OR:2010}%
  \BibitemOpen
  \bibfield  {author} {\bibinfo {author} {\bibfnamefont {J.-P.}\ \bibnamefont
  {Onnela}}\ and\ \bibinfo {author} {\bibfnamefont {F.}~\bibnamefont
  {Reed-Tsochas}},\ }\bibfield  {title} {\enquote {\bibinfo {title}
  {Spontaneous emergence of social influence in online systems},}\ }\href@noop
  {} {\bibfield  {journal} {\bibinfo  {journal} {Proc. Nat. Acad. Sci. (USA)}\
  }\textbf {\bibinfo {volume} {107}},\ \bibinfo {pages} {18375--18380}
  (\bibinfo {year} {2010})}\BibitemShut {NoStop}%
\bibitem [{\citenamefont {Romero}, \citenamefont {Meeder},\ and\ \citenamefont
  {Kleinberg}(2011)}]{RMK:2011}%
  \BibitemOpen
  \bibfield  {author} {\bibinfo {author} {\bibfnamefont {D.~M.}\ \bibnamefont
  {Romero}}, \bibinfo {author} {\bibfnamefont {B.}~\bibnamefont {Meeder}}, \
  and\ \bibinfo {author} {\bibfnamefont {J.}~\bibnamefont {Kleinberg}},\
  }\bibfield  {title} {\enquote {\bibinfo {title} {Differences in the mechanics
  of information diffusion across topics: idioms, political hashtags, and
  complex contagion on {Twitter}},}\ }in\ \href@noop {} {\emph {\bibinfo
  {booktitle} {Proceedings of the 20th International Conference on World Wide
  Web}}}\ (\bibinfo {organization} {ACM},\ \bibinfo {year} {2011})\ pp.\
  \bibinfo {pages} {695--704}\BibitemShut {NoStop}%
\bibitem [{\citenamefont {Yang}\ and\ \citenamefont
  {Leskovec}(2011)}]{YL:2011}%
  \BibitemOpen
  \bibfield  {author} {\bibinfo {author} {\bibfnamefont {J.}~\bibnamefont
  {Yang}}\ and\ \bibinfo {author} {\bibfnamefont {J.}~\bibnamefont
  {Leskovec}},\ }\bibfield  {title} {\enquote {\bibinfo {title} {Patterns of
  temporal variation in online media},}\ }in\ \href@noop {} {\emph {\bibinfo
  {booktitle} {Proceedings of the Fourth ACM International Conference on Web
  Search and Data Mining}}}\ (\bibinfo {organization} {ACM},\ \bibinfo {year}
  {2011})\ pp.\ \bibinfo {pages} {177--186}\BibitemShut {NoStop}%
\bibitem [{\citenamefont {Gon{\c{c}}alves}, \citenamefont {Perra},\ and\
  \citenamefont {Vespignani}(2011)}]{GPV:2011}%
  \BibitemOpen
  \bibfield  {author} {\bibinfo {author} {\bibfnamefont {B.}~\bibnamefont
  {Gon{\c{c}}alves}}, \bibinfo {author} {\bibfnamefont {N.}~\bibnamefont
  {Perra}}, \ and\ \bibinfo {author} {\bibfnamefont {A.}~\bibnamefont
  {Vespignani}},\ }\bibfield  {title} {\enquote {\bibinfo {title} {Modeling
  users' activity on {Twitter} networks: Validation of dunbar's number},}\
  }\href@noop {} {\bibfield  {journal} {\bibinfo  {journal} {PloS One}\
  }\textbf {\bibinfo {volume} {6}},\ \bibinfo {pages} {e22656} (\bibinfo {year}
  {2011})}\BibitemShut {NoStop}%
\bibitem [{\citenamefont {Weng}\ \emph {et~al.}(2012)\citenamefont {Weng},
  \citenamefont {Flammini}, \citenamefont {Vespignani},\ and\ \citenamefont
  {Menczer}}]{WFVM:2012}%
  \BibitemOpen
  \bibfield  {author} {\bibinfo {author} {\bibfnamefont {L.}~\bibnamefont
  {Weng}}, \bibinfo {author} {\bibfnamefont {A.}~\bibnamefont {Flammini}},
  \bibinfo {author} {\bibfnamefont {A.}~\bibnamefont {Vespignani}}, \ and\
  \bibinfo {author} {\bibfnamefont {F.}~\bibnamefont {Menczer}},\ }\bibfield
  {title} {\enquote {\bibinfo {title} {Competition among memes in a world with
  limited attention},}\ }\href@noop {} {\bibfield  {journal} {\bibinfo
  {journal} {Sci. Rep.}\ }\textbf {\bibinfo {volume} {2}},\ \bibinfo {pages}
  {335} (\bibinfo {year} {2012})}\BibitemShut {NoStop}%
\bibitem [{\citenamefont {Goel}, \citenamefont {Watts},\ and\ \citenamefont
  {Goldstein}(2012)}]{GWG:2012}%
  \BibitemOpen
  \bibfield  {author} {\bibinfo {author} {\bibfnamefont {S.}~\bibnamefont
  {Goel}}, \bibinfo {author} {\bibfnamefont {D.~J.}\ \bibnamefont {Watts}}, \
  and\ \bibinfo {author} {\bibfnamefont {D.~G.}\ \bibnamefont {Goldstein}},\
  }\bibfield  {title} {\enquote {\bibinfo {title} {The structure of online
  diffusion networks},}\ }in\ \href@noop {} {\emph {\bibinfo {booktitle}
  {Proceedings of the 13th ACM Conference on Electronic Commerce}}}\ (\bibinfo
  {organization} {ACM},\ \bibinfo {year} {2012})\ pp.\ \bibinfo {pages}
  {623--638}\BibitemShut {NoStop}%
\bibitem [{\citenamefont {Borondo}\ \emph {et~al.}(2012)\citenamefont
  {Borondo}, \citenamefont {Morales}, \citenamefont {Losada},\ and\
  \citenamefont {Benito}}]{BMLB:2012}%
  \BibitemOpen
  \bibfield  {author} {\bibinfo {author} {\bibfnamefont {J.}~\bibnamefont
  {Borondo}}, \bibinfo {author} {\bibfnamefont {A.~J.}\ \bibnamefont
  {Morales}}, \bibinfo {author} {\bibfnamefont {J.~C.}\ \bibnamefont {Losada}},
  \ and\ \bibinfo {author} {\bibfnamefont {R.~M.}\ \bibnamefont {Benito}},\
  }\bibfield  {title} {\enquote {\bibinfo {title} {Characterizing and modeling
  an electoral campaign in the context of twitter: 2011 spanish presidential
  election as a case study},}\ }\href@noop {} {\bibfield  {journal} {\bibinfo
  {journal} {Chaos}\ }\textbf {\bibinfo {volume} {22}},\ \bibinfo {pages}
  {023138} (\bibinfo {year} {2012})}\BibitemShut {NoStop}%
\bibitem [{\citenamefont {Gao}\ \emph {et~al.}(2012)\citenamefont {Gao},
  \citenamefont {Abel}, \citenamefont {Houben},\ and\ \citenamefont
  {Yu}}]{GAHY:2012}%
  \BibitemOpen
  \bibfield  {author} {\bibinfo {author} {\bibfnamefont {Q.}~\bibnamefont
  {Gao}}, \bibinfo {author} {\bibfnamefont {F.}~\bibnamefont {Abel}}, \bibinfo
  {author} {\bibfnamefont {G.-J.}\ \bibnamefont {Houben}}, \ and\ \bibinfo
  {author} {\bibfnamefont {Y.}~\bibnamefont {Yu}},\ }\bibfield  {title}
  {\enquote {\bibinfo {title} {A comparative study of users' microblogging
  behavior on {Sina Weibo and Twitter}},}\ }in\ \href@noop {} {\emph {\bibinfo
  {booktitle} {User Modeling, Adaptation, and Personalization}}}\ (\bibinfo
  {publisher} {Springer},\ \bibinfo {year} {2012})\ pp.\ \bibinfo {pages}
  {88--101}\BibitemShut {NoStop}%
\bibitem [{\citenamefont {Yu}, \citenamefont {Asur},\ and\ \citenamefont
  {Huberman}(2011)}]{YAH:2012}%
  \BibitemOpen
  \bibfield  {author} {\bibinfo {author} {\bibfnamefont {L.}~\bibnamefont
  {Yu}}, \bibinfo {author} {\bibfnamefont {S.}~\bibnamefont {Asur}}, \ and\
  \bibinfo {author} {\bibfnamefont {B.~A.}\ \bibnamefont {Huberman}},\
  }\bibfield  {title} {\enquote {\bibinfo {title} {What trends in chinese
  social media},}\ }\href@noop {} {\bibfield  {journal} {\bibinfo  {journal}
  {arXiv:1107.3522}\ } (\bibinfo {year} {2011})}\BibitemShut {NoStop}%
\bibitem [{\citenamefont {Oka}\ and\ \citenamefont {Ikegami}(2013)}]{OI:2013}%
  \BibitemOpen
  \bibfield  {author} {\bibinfo {author} {\bibfnamefont {M.}~\bibnamefont
  {Oka}}\ and\ \bibinfo {author} {\bibfnamefont {T.}~\bibnamefont {Ikegami}},\
  }\bibfield  {title} {\enquote {\bibinfo {title} {Exploring default mode and
  information flow on the web},}\ }\href@noop {} {\bibfield  {journal}
  {\bibinfo  {journal} {PloS One}\ }\textbf {\bibinfo {volume} {8}},\ \bibinfo
  {pages} {e60398} (\bibinfo {year} {2013})}\BibitemShut {NoStop}%
\bibitem [{\citenamefont {Zhao}\ \emph {et~al.}(2013)\citenamefont {Zhao},
  \citenamefont {Yang}, \citenamefont {Zhang}, \citenamefont {Zhou},
  \citenamefont {Huang},\ and\ \citenamefont {Lai}}]{Zhaoetal:2013}%
  \BibitemOpen
  \bibfield  {author} {\bibinfo {author} {\bibfnamefont {Z.-D.}\ \bibnamefont
  {Zhao}}, \bibinfo {author} {\bibfnamefont {Z.}~\bibnamefont {Yang}}, \bibinfo
  {author} {\bibfnamefont {Z.}~\bibnamefont {Zhang}}, \bibinfo {author}
  {\bibfnamefont {T.}~\bibnamefont {Zhou}}, \bibinfo {author} {\bibfnamefont
  {Z.-G.}\ \bibnamefont {Huang}}, \ and\ \bibinfo {author} {\bibfnamefont
  {Y.-C.}\ \bibnamefont {Lai}},\ }\bibfield  {title} {\enquote {\bibinfo
  {title} {Emergence of scaling in human-interest dynamics},}\ }\href@noop {}
  {\bibfield  {journal} {\bibinfo  {journal} {Sci. Rep.}\ }\textbf {\bibinfo
  {volume} {3}},\ \bibinfo {pages} {3472} (\bibinfo {year} {2013})}\BibitemShut
  {NoStop}%
\bibitem [{\citenamefont {Wang}\ \emph {et~al.}(2014)\citenamefont {Wang},
  \citenamefont {Huang}, \citenamefont {Rong}, \citenamefont {Wang},\ and\
  \citenamefont {Lai}}]{WHRWL:2014}%
  \BibitemOpen
  \bibfield  {author} {\bibinfo {author} {\bibfnamefont {L.-Z.}\ \bibnamefont
  {Wang}}, \bibinfo {author} {\bibfnamefont {Z.-G.}\ \bibnamefont {Huang}},
  \bibinfo {author} {\bibfnamefont {Z.-H.}\ \bibnamefont {Rong}}, \bibinfo
  {author} {\bibfnamefont {X.-F.}\ \bibnamefont {Wang}}, \ and\ \bibinfo
  {author} {\bibfnamefont {Y.-C.}\ \bibnamefont {Lai}},\ }\bibfield  {title}
  {\enquote {\bibinfo {title} {Emergence, evolution and scaling of online
  social networks},}\ }\href@noop {} {\bibfield  {journal} {\bibinfo  {journal}
  {PloS One}\ }\textbf {\bibinfo {volume} {9}},\ \bibinfo {pages} {e111013}
  (\bibinfo {year} {2014})}\BibitemShut {NoStop}%
\bibitem [{\citenamefont {Gleeson}\ \emph
  {et~al.}(2014{\natexlab{a}})\citenamefont {Gleeson}, \citenamefont {Ward},
  \citenamefont {O’Sullivan},\ and\ \citenamefont {Lee}}]{GWOL:2014}%
  \BibitemOpen
  \bibfield  {author} {\bibinfo {author} {\bibfnamefont {J.~P.}\ \bibnamefont
  {Gleeson}}, \bibinfo {author} {\bibfnamefont {J.~A.}\ \bibnamefont {Ward}},
  \bibinfo {author} {\bibfnamefont {K.~P.}\ \bibnamefont {O’Sullivan}}, \
  and\ \bibinfo {author} {\bibfnamefont {W.~T.}\ \bibnamefont {Lee}},\
  }\bibfield  {title} {\enquote {\bibinfo {title} {Competition-induced
  criticality in a model of meme popularity},}\ }\href@noop {} {\bibfield
  {journal} {\bibinfo  {journal} {Phys. Rev. Lett.}\ }\textbf {\bibinfo
  {volume} {112}},\ \bibinfo {pages} {048701} (\bibinfo {year}
  {2014}{\natexlab{a}})}\BibitemShut {NoStop}%
\bibitem [{\citenamefont {Lin}\ \emph {et~al.}(2014)\citenamefont {Lin},
  \citenamefont {Keegan}, \citenamefont {Margolin},\ and\ \citenamefont
  {Lazer}}]{LKML:2014}%
  \BibitemOpen
  \bibfield  {author} {\bibinfo {author} {\bibfnamefont {Y.-R.}\ \bibnamefont
  {Lin}}, \bibinfo {author} {\bibfnamefont {B.}~\bibnamefont {Keegan}},
  \bibinfo {author} {\bibfnamefont {D.}~\bibnamefont {Margolin}}, \ and\
  \bibinfo {author} {\bibfnamefont {D.}~\bibnamefont {Lazer}},\ }\bibfield
  {title} {\enquote {\bibinfo {title} {Rising tides or rising stars?: Dynamics
  of shared attention on {Twitter} during media events},}\ }\href@noop {}
  {\bibfield  {journal} {\bibinfo  {journal} {PLoS One}\ }\textbf {\bibinfo
  {volume} {9}},\ \bibinfo {pages} {e94093} (\bibinfo {year}
  {2014})}\BibitemShut {NoStop}%
\bibitem [{\citenamefont {Gleeson}\ \emph
  {et~al.}(2014{\natexlab{b}})\citenamefont {Gleeson}, \citenamefont {Cellai},
  \citenamefont {Onnela}, \citenamefont {Porter},\ and\ \citenamefont
  {Reed-Tsochas}}]{GCOPR:2014}%
  \BibitemOpen
  \bibfield  {author} {\bibinfo {author} {\bibfnamefont {J.~P.}\ \bibnamefont
  {Gleeson}}, \bibinfo {author} {\bibfnamefont {D.}~\bibnamefont {Cellai}},
  \bibinfo {author} {\bibfnamefont {J.-P.}\ \bibnamefont {Onnela}}, \bibinfo
  {author} {\bibfnamefont {M.~A.}\ \bibnamefont {Porter}}, \ and\ \bibinfo
  {author} {\bibfnamefont {F.}~\bibnamefont {Reed-Tsochas}},\ }\bibfield
  {title} {\enquote {\bibinfo {title} {A simple generative model of collective
  online behavior},}\ }\href@noop {} {\bibfield  {journal} {\bibinfo  {journal}
  {Proc. Nat. Acad. Sci. (USA)}\ }\textbf {\bibinfo {volume} {111}},\ \bibinfo
  {pages} {10411--10415} (\bibinfo {year} {2014}{\natexlab{b}})}\BibitemShut
  {NoStop}%
\bibitem [{\citenamefont {Kleineberg}\ and\ \citenamefont
  {Bogu{\~n}{\'a}}(2015)}]{KB:2015}%
  \BibitemOpen
  \bibfield  {author} {\bibinfo {author} {\bibfnamefont {K.-K.}\ \bibnamefont
  {Kleineberg}}\ and\ \bibinfo {author} {\bibfnamefont {M.}~\bibnamefont
  {Bogu{\~n}{\'a}}},\ }\bibfield  {title} {\enquote {\bibinfo {title} {Digital
  ecology: coexistence and domination among interacting networks},}\
  }\href@noop {} {\bibfield  {journal} {\bibinfo  {journal} {Sci. Rep.}\
  }\textbf {\bibinfo {volume} {5}},\ \bibinfo {pages} {10268} (\bibinfo {year}
  {2015})}\BibitemShut {NoStop}%
\bibitem [{\citenamefont {Gleeson}\ \emph {et~al.}(2016)\citenamefont
  {Gleeson}, \citenamefont {O’Sullivan}, \citenamefont {Ba{\~n}os},\ and\
  \citenamefont {Moreno}}]{GOBM:2016}%
  \BibitemOpen
  \bibfield  {author} {\bibinfo {author} {\bibfnamefont {J.~P.}\ \bibnamefont
  {Gleeson}}, \bibinfo {author} {\bibfnamefont {K.~P.}\ \bibnamefont
  {O’Sullivan}}, \bibinfo {author} {\bibfnamefont {R.~A.}\ \bibnamefont
  {Ba{\~n}os}}, \ and\ \bibinfo {author} {\bibfnamefont {Y.}~\bibnamefont
  {Moreno}},\ }\bibfield  {title} {\enquote {\bibinfo {title} {Effects of
  network structure, competition and memory time on social spreading
  phenomena},}\ }\href@noop {} {\bibfield  {journal} {\bibinfo  {journal}
  {Phys. Rev. X}\ }\textbf {\bibinfo {volume} {6}},\ \bibinfo {pages} {021019}
  (\bibinfo {year} {2016})}\BibitemShut {NoStop}%
\bibitem [{\citenamefont {Zhang}\ \emph {et~al.}(2016)\citenamefont {Zhang},
  \citenamefont {Liu}, \citenamefont {Zhan}, \citenamefont {Lu}, \citenamefont
  {Zhang},\ and\ \citenamefont {Zhang}}]{Zhangetal:2016}%
  \BibitemOpen
  \bibfield  {author} {\bibinfo {author} {\bibfnamefont {Z.-K.}\ \bibnamefont
  {Zhang}}, \bibinfo {author} {\bibfnamefont {C.}~\bibnamefont {Liu}}, \bibinfo
  {author} {\bibfnamefont {X.-X.}\ \bibnamefont {Zhan}}, \bibinfo {author}
  {\bibfnamefont {X.}~\bibnamefont {Lu}}, \bibinfo {author} {\bibfnamefont
  {C.-X.}\ \bibnamefont {Zhang}}, \ and\ \bibinfo {author} {\bibfnamefont
  {Y.-C.}\ \bibnamefont {Zhang}},\ }\bibfield  {title} {\enquote {\bibinfo
  {title} {Dynamics of information diffusion and its applications on complex
  networks},}\ }\href@noop {} {\bibfield  {journal} {\bibinfo  {journal} {Phys.
  Rep.}\ }\textbf {\bibinfo {volume} {651}},\ \bibinfo {pages} {1--34}
  (\bibinfo {year} {2016})}\BibitemShut {NoStop}%
\bibitem [{\citenamefont {Wang}\ \emph {et~al.}(2016)\citenamefont {Wang},
  \citenamefont {Wu}, \citenamefont {Zhang},\ and\ \citenamefont
  {Janssen}}]{WWZJ:2016}%
  \BibitemOpen
  \bibfield  {author} {\bibinfo {author} {\bibfnamefont {C.-J.}\ \bibnamefont
  {Wang}}, \bibinfo {author} {\bibfnamefont {L.}~\bibnamefont {Wu}}, \bibinfo
  {author} {\bibfnamefont {J.}~\bibnamefont {Zhang}}, \ and\ \bibinfo {author}
  {\bibfnamefont {M.~A.}\ \bibnamefont {Janssen}},\ }\bibfield  {title}
  {\enquote {\bibinfo {title} {The collective direction of attention
  diffusion},}\ }\href@noop {} {\bibfield  {journal} {\bibinfo  {journal} {Sci.
  Rep.}\ }\textbf {\bibinfo {volume} {6}},\ \bibinfo {pages} {34059} (\bibinfo
  {year} {2016})}\BibitemShut {NoStop}%
\bibitem [{\citenamefont {Qiu}\ \emph {et~al.}(2017)\citenamefont {Qiu},
  \citenamefont {Oliveira}, \citenamefont {Shirazi}, \citenamefont {Flammini},\
  and\ \citenamefont {Menczer}}]{QOSFM:2017}%
  \BibitemOpen
  \bibfield  {author} {\bibinfo {author} {\bibfnamefont {X.}~\bibnamefont
  {Qiu}}, \bibinfo {author} {\bibfnamefont {D.~F.}\ \bibnamefont {Oliveira}},
  \bibinfo {author} {\bibfnamefont {A.~S.}\ \bibnamefont {Shirazi}}, \bibinfo
  {author} {\bibfnamefont {A.}~\bibnamefont {Flammini}}, \ and\ \bibinfo
  {author} {\bibfnamefont {F.}~\bibnamefont {Menczer}},\ }\bibfield  {title}
  {\enquote {\bibinfo {title} {Limited individual attention and online virality
  of low-quality information},}\ }\href@noop {} {\bibfield  {journal} {\bibinfo
   {journal} {Nat. Human Behav.}\ }\textbf {\bibinfo {volume} {1}},\ \bibinfo
  {pages} {0132} (\bibinfo {year} {2017})}\BibitemShut {NoStop}%
\bibitem [{\citenamefont {Luo}\ \emph {et~al.}(2017)\citenamefont {Luo},
  \citenamefont {Morone}, \citenamefont {Sarraute}, \citenamefont {Travizano},\
  and\ \citenamefont {Makse}}]{LMSTM:2017}%
  \BibitemOpen
  \bibfield  {author} {\bibinfo {author} {\bibfnamefont {S.}~\bibnamefont
  {Luo}}, \bibinfo {author} {\bibfnamefont {F.}~\bibnamefont {Morone}},
  \bibinfo {author} {\bibfnamefont {C.}~\bibnamefont {Sarraute}}, \bibinfo
  {author} {\bibfnamefont {M.}~\bibnamefont {Travizano}}, \ and\ \bibinfo
  {author} {\bibfnamefont {H.~A.}\ \bibnamefont {Makse}},\ }\bibfield  {title}
  {\enquote {\bibinfo {title} {Inferring personal economic status from social
  network location},}\ }\href@noop {} {\bibfield  {journal} {\bibinfo
  {journal} {Nat. Commun.}\ }\textbf {\bibinfo {volume} {8}},\ \bibinfo {pages}
  {15227} (\bibinfo {year} {2017})}\BibitemShut {NoStop}%
\bibitem [{\citenamefont {Lehmann}\ \emph {et~al.}(2012)\citenamefont
  {Lehmann}, \citenamefont {Gonçalves}, \citenamefont {Ramasco},\ and\
  \citenamefont {Cattuto}}]{LGR:2012}%
  \BibitemOpen
  \bibfield  {author} {\bibinfo {author} {\bibfnamefont {J.}~\bibnamefont
  {Lehmann}}, \bibinfo {author} {\bibfnamefont {B.}~\bibnamefont {Gonçalves}},
  \bibinfo {author} {\bibfnamefont {J.~J.}\ \bibnamefont {Ramasco}}, \ and\
  \bibinfo {author} {\bibfnamefont {C.}~\bibnamefont {Cattuto}},\ }\bibfield
  {title} {\enquote {\bibinfo {title} {Dynamical classes of collective
  attention in twitter},}\ }in\ \href@noop {} {\emph {\bibinfo {booktitle}
  {WWW'12 - Proceedings of the 21st international conference on World Wide
  Web}}}\ (\bibinfo  {publisher} {ACM},\ \bibinfo {address} {New York},\
  \bibinfo {year} {2012})\ pp.\ \bibinfo {pages} {251--260}\BibitemShut
  {NoStop}%
\bibitem [{\citenamefont {Kleineberg}\ and\ \citenamefont
  {Bogun{\'a}}(2014)}]{KB:2014}%
  \BibitemOpen
  \bibfield  {author} {\bibinfo {author} {\bibfnamefont {K.-K.}\ \bibnamefont
  {Kleineberg}}\ and\ \bibinfo {author} {\bibfnamefont {M.}~\bibnamefont
  {Bogun{\'a}}},\ }\bibfield  {title} {\enquote {\bibinfo {title} {Evolution of
  the digital society reveals balance between viral and mass media
  influence},}\ }\href@noop {} {\bibfield  {journal} {\bibinfo  {journal}
  {Phys. Rev. X}\ }\textbf {\bibinfo {volume} {4}},\ \bibinfo {pages} {031046}
  (\bibinfo {year} {2014})}\BibitemShut {NoStop}%
\bibitem [{\citenamefont {Zaman}\ \emph {et~al.}(2014)\citenamefont {Zaman},
  \citenamefont {Fox}, \citenamefont {Bradlow} \emph {et~al.}}]{ZFB:2014}%
  \BibitemOpen
  \bibfield  {author} {\bibinfo {author} {\bibfnamefont {T.}~\bibnamefont
  {Zaman}}, \bibinfo {author} {\bibfnamefont {E.~B.}\ \bibnamefont {Fox}},
  \bibinfo {author} {\bibfnamefont {E.~T.}\ \bibnamefont {Bradlow}},  \emph
  {et~al.},\ }\bibfield  {title} {\enquote {\bibinfo {title} {A {Bayesian}
  approach for predicting the popularity of tweets},}\ }\href@noop {}
  {\bibfield  {journal} {\bibinfo  {journal} {Ann. Appl. Stat.}\ }\textbf
  {\bibinfo {volume} {8}},\ \bibinfo {pages} {1583--1611} (\bibinfo {year}
  {2014})}\BibitemShut {NoStop}%
\bibitem [{\citenamefont {Borge-Holthoefer}\ \emph {et~al.}(2017)\citenamefont
  {Borge-Holthoefer}, \citenamefont {Banos}, \citenamefont {Gracia-Lazaro},\
  and\ \citenamefont {Moreno}}]{BHBGLM:2017}%
  \BibitemOpen
  \bibfield  {author} {\bibinfo {author} {\bibfnamefont {J.}~\bibnamefont
  {Borge-Holthoefer}}, \bibinfo {author} {\bibfnamefont {R.~A.}\ \bibnamefont
  {Banos}}, \bibinfo {author} {\bibfnamefont {C.}~\bibnamefont
  {Gracia-Lazaro}}, \ and\ \bibinfo {author} {\bibfnamefont {Y.}~\bibnamefont
  {Moreno}},\ }\bibfield  {title} {\enquote {\bibinfo {title} {Emergence of
  consensus as a modular-to-nested transition in communication dynamics},}\
  }\href@noop {} {\bibfield  {journal} {\bibinfo  {journal} {Sci. Rep.}\
  }\textbf {\bibinfo {volume} {7}},\ \bibinfo {pages} {41673} (\bibinfo {year}
  {2017})}\BibitemShut {NoStop}%
\bibitem [{\citenamefont {Hawkes}(1971)}]{Hawkes:1971}%
  \BibitemOpen
  \bibfield  {author} {\bibinfo {author} {\bibfnamefont {A.~G.}\ \bibnamefont
  {Hawkes}},\ }\bibfield  {title} {\enquote {\bibinfo {title} {Spectra of some
  self-exciting and mutually exciting point processes},}\ }\href@noop {}
  {\bibfield  {journal} {\bibinfo  {journal} {Biometrika}\ }\textbf {\bibinfo
  {volume} {58}},\ \bibinfo {pages} {83--90} (\bibinfo {year}
  {1971})}\BibitemShut {NoStop}%
\bibitem [{\citenamefont {Zhao}\ \emph {et~al.}(2015)\citenamefont {Zhao},
  \citenamefont {Erdogdu}, \citenamefont {He}, \citenamefont {Rajaraman},\ and\
  \citenamefont {Leskovec}}]{ZEHRL:2015}%
  \BibitemOpen
  \bibfield  {author} {\bibinfo {author} {\bibfnamefont {Q.}~\bibnamefont
  {Zhao}}, \bibinfo {author} {\bibfnamefont {M.~A.}\ \bibnamefont {Erdogdu}},
  \bibinfo {author} {\bibfnamefont {H.~Y.}\ \bibnamefont {He}}, \bibinfo
  {author} {\bibfnamefont {A.}~\bibnamefont {Rajaraman}}, \ and\ \bibinfo
  {author} {\bibfnamefont {J.}~\bibnamefont {Leskovec}},\ }\bibfield  {title}
  {\enquote {\bibinfo {title} {Seismic: A self-exciting point process model for
  predicting tweet popularity},}\ }in\ \href@noop {} {\emph {\bibinfo
  {booktitle} {Proceedings of the 21th ACM SIGKDD International Conference on
  Knowledge Discovery and Data Mining}}}\ (\bibinfo {organization} {ACM},\
  \bibinfo {year} {2015})\ pp.\ \bibinfo {pages} {1513--1522}\BibitemShut
  {NoStop}%
\bibitem [{\citenamefont {Pastor-Satorras}\ and\ \citenamefont
  {Vespignani}(2001)}]{PSRV:2001}%
  \BibitemOpen
  \bibfield  {author} {\bibinfo {author} {\bibfnamefont {R.}~\bibnamefont
  {Pastor-Satorras}}\ and\ \bibinfo {author} {\bibfnamefont {A.}~\bibnamefont
  {Vespignani}},\ }\bibfield  {title} {\enquote {\bibinfo {title} {Epidemic
  spreading in scale-free networks},}\ }\href@noop {} {\bibfield  {journal}
  {\bibinfo  {journal} {Phys. Rev. Lett.}\ }\textbf {\bibinfo {volume} {86}},\
  \bibinfo {pages} {3200} (\bibinfo {year} {2001})}\BibitemShut {NoStop}%
\bibitem [{\citenamefont {Newman}(2002)}]{Newman:2002}%
  \BibitemOpen
  \bibfield  {author} {\bibinfo {author} {\bibfnamefont {M.~E.~J.}\
  \bibnamefont {Newman}},\ }\bibfield  {title} {\enquote {\bibinfo {title}
  {Spread of epidemic disease on networks},}\ }\href@noop {} {\bibfield
  {journal} {\bibinfo  {journal} {Phys. Rev. E}\ }\textbf {\bibinfo {volume}
  {66}},\ \bibinfo {pages} {016128} (\bibinfo {year} {2002})}\BibitemShut
  {NoStop}%
\bibitem [{\citenamefont {Pastor-Satorras}\ \emph {et~al.}(2015)\citenamefont
  {Pastor-Satorras}, \citenamefont {Castellano}, \citenamefont {Van~Mieghem},\
  and\ \citenamefont {Vespignani}}]{PCVV:2015}%
  \BibitemOpen
  \bibfield  {author} {\bibinfo {author} {\bibfnamefont {R.}~\bibnamefont
  {Pastor-Satorras}}, \bibinfo {author} {\bibfnamefont {C.}~\bibnamefont
  {Castellano}}, \bibinfo {author} {\bibfnamefont {P.}~\bibnamefont
  {Van~Mieghem}}, \ and\ \bibinfo {author} {\bibfnamefont {A.}~\bibnamefont
  {Vespignani}},\ }\bibfield  {title} {\enquote {\bibinfo {title} {Epidemic
  processes in complex networks},}\ }\href@noop {} {\bibfield  {journal}
  {\bibinfo  {journal} {Rev. Mod. Phys.}\ }\textbf {\bibinfo {volume} {87}},\
  \bibinfo {pages} {925} (\bibinfo {year} {2015})}\BibitemShut {NoStop}%
\bibitem [{\citenamefont {von Foerster}(1959)}]{von:1959}%
  \BibitemOpen
  \bibfield  {author} {\bibinfo {author} {\bibfnamefont {H.}~\bibnamefont {von
  Foerster}},\ }\bibfield  {title} {\enquote {\bibinfo {title} {Some remarks on
  changing populations},}\ }\href@noop {} {\bibfield  {journal} {\bibinfo
  {journal} {Kinet. Cellu. Prolif.}\ ,\ \bibinfo {pages} {382--407}} (\bibinfo
  {year} {1959})}\BibitemShut {NoStop}%
\bibitem [{\citenamefont {Trucco}(1965{\natexlab{a}})}]{Trucco1:1965}%
  \BibitemOpen
  \bibfield  {author} {\bibinfo {author} {\bibfnamefont {E.}~\bibnamefont
  {Trucco}},\ }\bibfield  {title} {\enquote {\bibinfo {title} {Mathematical
  models for cellular systems the von {Foerster} equation. part i},}\
  }\href@noop {} {\bibfield  {journal} {\bibinfo  {journal} {Bull. Math.
  Biol.}\ }\textbf {\bibinfo {volume} {27}},\ \bibinfo {pages} {285--304}
  (\bibinfo {year} {1965}{\natexlab{a}})}\BibitemShut {NoStop}%
\bibitem [{\citenamefont {Trucco}(1965{\natexlab{b}})}]{Trucco2:1965}%
  \BibitemOpen
  \bibfield  {author} {\bibinfo {author} {\bibfnamefont {E.}~\bibnamefont
  {Trucco}},\ }\bibfield  {title} {\enquote {\bibinfo {title} {Mathematical
  models for cellular systems. the von {Foerster} equation. part ii},}\
  }\href@noop {} {\bibfield  {journal} {\bibinfo  {journal} {Bull. Math.
  Biol.}\ }\textbf {\bibinfo {volume} {27}},\ \bibinfo {pages} {449--471}
  (\bibinfo {year} {1965}{\natexlab{b}})}\BibitemShut {NoStop}%
\bibitem [{\citenamefont {Lu}\ \emph {et~al.}(2004)\citenamefont {Lu},
  \citenamefont {Volfson}, \citenamefont {Tsimring},\ and\ \citenamefont
  {Hasty}}]{LVTH:2004}%
  \BibitemOpen
  \bibfield  {author} {\bibinfo {author} {\bibfnamefont {T.}~\bibnamefont
  {Lu}}, \bibinfo {author} {\bibfnamefont {D.}~\bibnamefont {Volfson}},
  \bibinfo {author} {\bibfnamefont {L.}~\bibnamefont {Tsimring}}, \ and\
  \bibinfo {author} {\bibfnamefont {J.}~\bibnamefont {Hasty}},\ }\bibfield
  {title} {\enquote {\bibinfo {title} {Cellular growth and division in the
  {Gillespie} algorithm},}\ }\href@noop {} {\bibfield  {journal} {\bibinfo
  {journal} {Sys. Biol.}\ }\textbf {\bibinfo {volume} {1}},\ \bibinfo {pages}
  {121--128} (\bibinfo {year} {2004})}\BibitemShut {NoStop}%
\bibitem [{\citenamefont {Kutalik}, \citenamefont {Razaz},\ and\ \citenamefont
  {Baranyi}(2005)}]{KRB:2005}%
  \BibitemOpen
  \bibfield  {author} {\bibinfo {author} {\bibfnamefont {Z.}~\bibnamefont
  {Kutalik}}, \bibinfo {author} {\bibfnamefont {M.}~\bibnamefont {Razaz}}, \
  and\ \bibinfo {author} {\bibfnamefont {J.}~\bibnamefont {Baranyi}},\
  }\bibfield  {title} {\enquote {\bibinfo {title} {Connection between
  stochastic and deterministic modelling of microbial growth},}\ }\href@noop {}
  {\bibfield  {journal} {\bibinfo  {journal} {J. Theo. Biol.}\ }\textbf
  {\bibinfo {volume} {232}},\ \bibinfo {pages} {285--299} (\bibinfo {year}
  {2005})}\BibitemShut {NoStop}%
\bibitem [{\citenamefont {Wang}\ \emph {et~al.}(2010)\citenamefont {Wang},
  \citenamefont {Robert}, \citenamefont {Pelletier}, \citenamefont {Dang},
  \citenamefont {Taddei}, \citenamefont {Wright},\ and\ \citenamefont
  {Jun}}]{Wangetal:2010}%
  \BibitemOpen
  \bibfield  {author} {\bibinfo {author} {\bibfnamefont {P.}~\bibnamefont
  {Wang}}, \bibinfo {author} {\bibfnamefont {L.}~\bibnamefont {Robert}},
  \bibinfo {author} {\bibfnamefont {J.}~\bibnamefont {Pelletier}}, \bibinfo
  {author} {\bibfnamefont {W.~L.}\ \bibnamefont {Dang}}, \bibinfo {author}
  {\bibfnamefont {F.}~\bibnamefont {Taddei}}, \bibinfo {author} {\bibfnamefont
  {A.}~\bibnamefont {Wright}}, \ and\ \bibinfo {author} {\bibfnamefont
  {S.}~\bibnamefont {Jun}},\ }\bibfield  {title} {\enquote {\bibinfo {title}
  {Robust growth of {Escherichia} coli},}\ }\href@noop {} {\bibfield  {journal}
  {\bibinfo  {journal} {Curr. Biol.}\ }\textbf {\bibinfo {volume} {20}},\
  \bibinfo {pages} {1099--1103} (\bibinfo {year} {2010})}\BibitemShut {NoStop}%
\bibitem [{\citenamefont {Horowitz}\ \emph {et~al.}(2010)\citenamefont
  {Horowitz}, \citenamefont {Normand}, \citenamefont {Corradini},\ and\
  \citenamefont {Peleg}}]{Horowitz:2010}%
  \BibitemOpen
  \bibfield  {author} {\bibinfo {author} {\bibfnamefont {J.}~\bibnamefont
  {Horowitz}}, \bibinfo {author} {\bibfnamefont {M.~D.}\ \bibnamefont
  {Normand}}, \bibinfo {author} {\bibfnamefont {M.~G.}\ \bibnamefont
  {Corradini}}, \ and\ \bibinfo {author} {\bibfnamefont {M.}~\bibnamefont
  {Peleg}},\ }\bibfield  {title} {\enquote {\bibinfo {title} {Probabilistic
  model of microbial cell growth, division, and mortality},}\ }\href@noop {}
  {\bibfield  {journal} {\bibinfo  {journal} {Appl. Envir. Microbiol.}\
  }\textbf {\bibinfo {volume} {76}},\ \bibinfo {pages} {230--242} (\bibinfo
  {year} {2010})}\BibitemShut {NoStop}%
\bibitem [{\citenamefont {Stukalin}\ \emph {et~al.}(2013)\citenamefont
  {Stukalin}, \citenamefont {Aifuwa}, \citenamefont {Kim}, \citenamefont
  {Wirtz},\ and\ \citenamefont {Sun}}]{SAKWS:2013}%
  \BibitemOpen
  \bibfield  {author} {\bibinfo {author} {\bibfnamefont {E.~B.}\ \bibnamefont
  {Stukalin}}, \bibinfo {author} {\bibfnamefont {I.}~\bibnamefont {Aifuwa}},
  \bibinfo {author} {\bibfnamefont {J.~S.}\ \bibnamefont {Kim}}, \bibinfo
  {author} {\bibfnamefont {D.}~\bibnamefont {Wirtz}}, \ and\ \bibinfo {author}
  {\bibfnamefont {S.~X.}\ \bibnamefont {Sun}},\ }\bibfield  {title} {\enquote
  {\bibinfo {title} {Age-dependent stochastic models for understanding
  population fluctuations in continuously cultured cells},}\ }\href@noop {}
  {\bibfield  {journal} {\bibinfo  {journal} {J. Roy. Soc. Interface}\ }\textbf
  {\bibinfo {volume} {10}},\ \bibinfo {pages} {20130325} (\bibinfo {year}
  {2013})}\BibitemShut {NoStop}%
\bibitem [{\citenamefont {Ribeiro}(2014)}]{Ribeiro:2014}%
  \BibitemOpen
  \bibfield  {author} {\bibinfo {author} {\bibfnamefont {B.}~\bibnamefont
  {Ribeiro}},\ }\bibfield  {title} {\enquote {\bibinfo {title} {Modeling and
  predicting the growth and death of membership-based websites},}\ }in\
  \href@noop {} {\emph {\bibinfo {booktitle} {Proceedings of the 23rd
  International Conference on World Wide Web}}}\ (\bibinfo {organization}
  {ACM},\ \bibinfo {year} {2014})\ pp.\ \bibinfo {pages} {653--664}\BibitemShut
  {NoStop}%
\bibitem [{\citenamefont {Antunes}\ and\ \citenamefont
  {Singh}(2015)}]{AS:2015}%
  \BibitemOpen
  \bibfield  {author} {\bibinfo {author} {\bibfnamefont {D.}~\bibnamefont
  {Antunes}}\ and\ \bibinfo {author} {\bibfnamefont {A.}~\bibnamefont
  {Singh}},\ }\bibfield  {title} {\enquote {\bibinfo {title} {Quantifying gene
  expression variability arising from randomness in cell division times},}\
  }\href@noop {} {\bibfield  {journal} {\bibinfo  {journal} {J. Math. Biol.}\
  }\textbf {\bibinfo {volume} {71}},\ \bibinfo {pages} {437--463} (\bibinfo
  {year} {2015})}\BibitemShut {NoStop}%
\bibitem [{\citenamefont {Greenman}\ and\ \citenamefont
  {Chou}(2016)}]{GC:2016}%
  \BibitemOpen
  \bibfield  {author} {\bibinfo {author} {\bibfnamefont {C.~D.}\ \bibnamefont
  {Greenman}}\ and\ \bibinfo {author} {\bibfnamefont {T.}~\bibnamefont
  {Chou}},\ }\bibfield  {title} {\enquote {\bibinfo {title} {Kinetic theory of
  age-structured stochastic birth-death processes},}\ }\href@noop {} {\bibfield
   {journal} {\bibinfo  {journal} {Phys. Rev. E}\ }\textbf {\bibinfo {volume}
  {93}},\ \bibinfo {pages} {012112} (\bibinfo {year} {2016})}\BibitemShut
  {NoStop}%
\bibitem [{\citenamefont {Chou}\ and\ \citenamefont
  {Greenman}(2016)}]{CG:2016}%
  \BibitemOpen
  \bibfield  {author} {\bibinfo {author} {\bibfnamefont {T.}~\bibnamefont
  {Chou}}\ and\ \bibinfo {author} {\bibfnamefont {C.~D.}\ \bibnamefont
  {Greenman}},\ }\bibfield  {title} {\enquote {\bibinfo {title} {A hierarchical
  kinetic theory of birth, death and fission in age-structured interacting
  populations},}\ }\href@noop {} {\bibfield  {journal} {\bibinfo  {journal} {J.
  Stat. Phys.}\ }\textbf {\bibinfo {volume} {164}},\ \bibinfo {pages} {49--76}
  (\bibinfo {year} {2016})}\BibitemShut {NoStop}%
\bibitem [{\citenamefont {Widder}\ \emph {et~al.}(2016)\citenamefont {Widder},
  \citenamefont {Allen}, \citenamefont {Pfeiffer}, \citenamefont {Curtis},
  \citenamefont {Wiuf}, \citenamefont {Sloan}, \citenamefont {Cordero},
  \citenamefont {Brown}, \citenamefont {Momeni}, \citenamefont {Shou} \emph
  {et~al.}}]{Widder:2016}%
  \BibitemOpen
  \bibfield  {author} {\bibinfo {author} {\bibfnamefont {S.}~\bibnamefont
  {Widder}}, \bibinfo {author} {\bibfnamefont {R.~J.}\ \bibnamefont {Allen}},
  \bibinfo {author} {\bibfnamefont {T.}~\bibnamefont {Pfeiffer}}, \bibinfo
  {author} {\bibfnamefont {T.~P.}\ \bibnamefont {Curtis}}, \bibinfo {author}
  {\bibfnamefont {C.}~\bibnamefont {Wiuf}}, \bibinfo {author} {\bibfnamefont
  {W.~T.}\ \bibnamefont {Sloan}}, \bibinfo {author} {\bibfnamefont {O.~X.}\
  \bibnamefont {Cordero}}, \bibinfo {author} {\bibfnamefont {S.~P.}\
  \bibnamefont {Brown}}, \bibinfo {author} {\bibfnamefont {B.}~\bibnamefont
  {Momeni}}, \bibinfo {author} {\bibfnamefont {W.}~\bibnamefont {Shou}},  \emph
  {et~al.},\ }\bibfield  {title} {\enquote {\bibinfo {title} {Challenges in
  microbial ecology: building predictive understanding of community function
  and dynamics},}\ }\href@noop {} {\bibfield  {journal} {\bibinfo  {journal}
  {ISME J.}\ }\textbf {\bibinfo {volume} {10}},\ \bibinfo {pages} {2557}
  (\bibinfo {year} {2016})}\BibitemShut {NoStop}%
\bibitem [{\citenamefont {Wetzker}, \citenamefont {Zimmermann},\ and\
  \citenamefont {Bauckhage}(2008)}]{WZB:2008}%
  \BibitemOpen
  \bibfield  {author} {\bibinfo {author} {\bibfnamefont {R.}~\bibnamefont
  {Wetzker}}, \bibinfo {author} {\bibfnamefont {C.}~\bibnamefont {Zimmermann}},
  \ and\ \bibinfo {author} {\bibfnamefont {C.}~\bibnamefont {Bauckhage}},\
  }\bibfield  {title} {\enquote {\bibinfo {title} {Analyzing social bookmarking
  systems: {A} del. icio. us cookbook},}\ }in\ \href@noop {} {\emph {\bibinfo
  {booktitle} {Proceedings of the ECAI 2008 Mining Social Data Workshop}}}\
  (\bibinfo {year} {2008})\ pp.\ \bibinfo {pages} {26--30}\BibitemShut
  {NoStop}%
\bibitem [{\citenamefont {Wikipedia}(2015)}]{weibo:2015}%
  \BibitemOpen
  \bibfield  {author} {\bibinfo {author} {\bibnamefont {Wikipedia}},\
  }\href@noop {} {\enquote {\bibinfo {title} {{Sina Weibo}},}\ }\bibinfo
  {howpublished} {\url{https://en.wikipedia.org/wiki/Sina_Weibo}} (\bibinfo
  {year} {2015})\BibitemShut {NoStop}%
\bibitem [{\citenamefont {Boyd}, \citenamefont {Golder},\ and\ \citenamefont
  {Lotan}(2010)}]{BGL:2010}%
  \BibitemOpen
  \bibfield  {author} {\bibinfo {author} {\bibfnamefont {D.}~\bibnamefont
  {Boyd}}, \bibinfo {author} {\bibfnamefont {S.}~\bibnamefont {Golder}}, \ and\
  \bibinfo {author} {\bibfnamefont {G.}~\bibnamefont {Lotan}},\ }\bibfield
  {title} {\enquote {\bibinfo {title} {Tweet, tweet, retweet: Conversational
  aspects of retweeting on {Twitter}},}\ }in\ \href@noop {} {\emph {\bibinfo
  {booktitle} {System Sciences (HICSS), 2010 43rd Hawaii International
  Conference on}}}\ (\bibinfo {organization} {IEEE},\ \bibinfo {year} {2010})\
  pp.\ \bibinfo {pages} {1--10}\BibitemShut {NoStop}%
\bibitem [{\citenamefont {Corder}\ and\ \citenamefont
  {Foreman}(2009)}]{CF:2009}%
  \BibitemOpen
  \bibfield  {author} {\bibinfo {author} {\bibfnamefont {G.~W.}\ \bibnamefont
  {Corder}}\ and\ \bibinfo {author} {\bibfnamefont {D.~I.}\ \bibnamefont
  {Foreman}},\ }\href@noop {} {\emph {\bibinfo {title} {Nonparametric
  Statistics for Non‐Statisticians: A Step-by-Step Approach}}}\ (\bibinfo
  {publisher} {John Wiley \& Sons},\ \bibinfo {year} {2009})\BibitemShut
  {NoStop}%
\bibitem [{\citenamefont {Barab{\'a}si}(2005)}]{Barabasi:2005}%
  \BibitemOpen
  \bibfield  {author} {\bibinfo {author} {\bibfnamefont {A.-L.}\ \bibnamefont
  {Barab{\'a}si}},\ }\bibfield  {title} {\enquote {\bibinfo {title} {The origin
  of bursts and heavy tails in human dynamics},}\ }\href@noop {} {\bibfield
  {journal} {\bibinfo  {journal} {Nature}\ }\textbf {\bibinfo {volume} {435}},\
  \bibinfo {pages} {207--211} (\bibinfo {year} {2005})}\BibitemShut {NoStop}%
\bibitem [{\citenamefont {Oliveira}\ and\ \citenamefont
  {Barab{\'a}si}(2005)}]{OB:2005}%
  \BibitemOpen
  \bibfield  {author} {\bibinfo {author} {\bibfnamefont {J.~G.}\ \bibnamefont
  {Oliveira}}\ and\ \bibinfo {author} {\bibfnamefont {A.-L.}\ \bibnamefont
  {Barab{\'a}si}},\ }\bibfield  {title} {\enquote {\bibinfo {title} {Human
  dynamics: {Darwin and Einstein} correspondence patterns},}\ }\href@noop {}
  {\bibfield  {journal} {\bibinfo  {journal} {Nature}\ }\textbf {\bibinfo
  {volume} {437}},\ \bibinfo {pages} {1251--1251} (\bibinfo {year}
  {2005})}\BibitemShut {NoStop}%
\bibitem [{\citenamefont {Dezs{\"o}}\ \emph {et~al.}(2006)\citenamefont
  {Dezs{\"o}}, \citenamefont {Almaas}, \citenamefont {Luk{\'a}cs},
  \citenamefont {R{\'a}cz}, \citenamefont {Szakad{\'a}t},\ and\ \citenamefont
  {Barab{\'a}si}}]{Dezso:2006}%
  \BibitemOpen
  \bibfield  {author} {\bibinfo {author} {\bibfnamefont {Z.}~\bibnamefont
  {Dezs{\"o}}}, \bibinfo {author} {\bibfnamefont {E.}~\bibnamefont {Almaas}},
  \bibinfo {author} {\bibfnamefont {A.}~\bibnamefont {Luk{\'a}cs}}, \bibinfo
  {author} {\bibfnamefont {B.}~\bibnamefont {R{\'a}cz}}, \bibinfo {author}
  {\bibfnamefont {I.}~\bibnamefont {Szakad{\'a}t}}, \ and\ \bibinfo {author}
  {\bibfnamefont {A.-L.}\ \bibnamefont {Barab{\'a}si}},\ }\bibfield  {title}
  {\enquote {\bibinfo {title} {Dynamics of information access on the web},}\
  }\href@noop {} {\bibfield  {journal} {\bibinfo  {journal} {Phys. Rev. E}\
  }\textbf {\bibinfo {volume} {73}},\ \bibinfo {pages} {066132} (\bibinfo
  {year} {2006})}\BibitemShut {NoStop}%
\bibitem [{\citenamefont {Zhou}\ \emph {et~al.}(2008)\citenamefont {Zhou},
  \citenamefont {Kiet}, \citenamefont {Kim}, \citenamefont {Wang},\ and\
  \citenamefont {Holme}}]{ZKKWH:2008}%
  \BibitemOpen
  \bibfield  {author} {\bibinfo {author} {\bibfnamefont {T.}~\bibnamefont
  {Zhou}}, \bibinfo {author} {\bibfnamefont {H.~A.-T.}\ \bibnamefont {Kiet}},
  \bibinfo {author} {\bibfnamefont {B.~J.}\ \bibnamefont {Kim}}, \bibinfo
  {author} {\bibfnamefont {B.-H.}\ \bibnamefont {Wang}}, \ and\ \bibinfo
  {author} {\bibfnamefont {P.}~\bibnamefont {Holme}},\ }\bibfield  {title}
  {\enquote {\bibinfo {title} {Role of activity in human dynamics},}\
  }\href@noop {} {\bibfield  {journal} {\bibinfo  {journal} {EPL (Europhys.
  Lett.)}\ }\textbf {\bibinfo {volume} {82}},\ \bibinfo {pages} {28002}
  (\bibinfo {year} {2008})}\BibitemShut {NoStop}%
\bibitem [{\citenamefont {Gon{\c{c}}alves}\ and\ \citenamefont
  {Ramasco}(2008)}]{GR:2008}%
  \BibitemOpen
  \bibfield  {author} {\bibinfo {author} {\bibfnamefont {B.}~\bibnamefont
  {Gon{\c{c}}alves}}\ and\ \bibinfo {author} {\bibfnamefont {J.~J.}\
  \bibnamefont {Ramasco}},\ }\bibfield  {title} {\enquote {\bibinfo {title}
  {Human dynamics revealed through {Web} analytics},}\ }\href@noop {}
  {\bibfield  {journal} {\bibinfo  {journal} {Phys. Rev. E}\ }\textbf {\bibinfo
  {volume} {78}},\ \bibinfo {pages} {026123} (\bibinfo {year}
  {2008})}\BibitemShut {NoStop}%
\bibitem [{\citenamefont {Song}\ \emph {et~al.}(2010)\citenamefont {Song},
  \citenamefont {Koren}, \citenamefont {Wang},\ and\ \citenamefont
  {Barab{\'a}si}}]{SKWB:2010}%
  \BibitemOpen
  \bibfield  {author} {\bibinfo {author} {\bibfnamefont {C.}~\bibnamefont
  {Song}}, \bibinfo {author} {\bibfnamefont {T.}~\bibnamefont {Koren}},
  \bibinfo {author} {\bibfnamefont {P.}~\bibnamefont {Wang}}, \ and\ \bibinfo
  {author} {\bibfnamefont {A.-L.}\ \bibnamefont {Barab{\'a}si}},\ }\bibfield
  {title} {\enquote {\bibinfo {title} {Modeling the scaling properties of human
  mobility},}\ }\href@noop {} {\bibfield  {journal} {\bibinfo  {journal} {Nat.
  Phys.}\ }\textbf {\bibinfo {volume} {6}},\ \bibinfo {pages} {818--823}
  (\bibinfo {year} {2010})}\BibitemShut {NoStop}%
\bibitem [{\citenamefont {Szell}\ \emph {et~al.}(2012)\citenamefont {Szell},
  \citenamefont {Sinatra}, \citenamefont {Petri}, \citenamefont {Thurner},\
  and\ \citenamefont {Latora}}]{SSPTL:2012}%
  \BibitemOpen
  \bibfield  {author} {\bibinfo {author} {\bibfnamefont {M.}~\bibnamefont
  {Szell}}, \bibinfo {author} {\bibfnamefont {R.}~\bibnamefont {Sinatra}},
  \bibinfo {author} {\bibfnamefont {G.}~\bibnamefont {Petri}}, \bibinfo
  {author} {\bibfnamefont {S.}~\bibnamefont {Thurner}}, \ and\ \bibinfo
  {author} {\bibfnamefont {V.}~\bibnamefont {Latora}},\ }\bibfield  {title}
  {\enquote {\bibinfo {title} {Understanding mobility in a social petri
  dish},}\ }\href@noop {} {\bibfield  {journal} {\bibinfo  {journal} {Sci.
  Rep.}\ }\textbf {\bibinfo {volume} {2}},\ \bibinfo {pages} {457} (\bibinfo
  {year} {2012})}\BibitemShut {NoStop}%
\bibitem [{\citenamefont {Wu}\ and\ \citenamefont {Huberman}(2007)}]{WH:2007}%
  \BibitemOpen
  \bibfield  {author} {\bibinfo {author} {\bibfnamefont {F.}~\bibnamefont
  {Wu}}\ and\ \bibinfo {author} {\bibfnamefont {B.~A.}\ \bibnamefont
  {Huberman}},\ }\bibfield  {title} {\enquote {\bibinfo {title} {Novelty and
  collective attention},}\ }\href@noop {} {\bibfield  {journal} {\bibinfo
  {journal} {Proc. Natl. Acad. Sci. (USA)}\ }\textbf {\bibinfo {volume}
  {104}},\ \bibinfo {pages} {17599--17601} (\bibinfo {year}
  {2007})}\BibitemShut {NoStop}%
\bibitem [{\citenamefont {Ye}\ \emph {et~al.}(2012)\citenamefont {Ye},
  \citenamefont {Sandholm}, \citenamefont {Wang}, \citenamefont {Aperjis},\
  and\ \citenamefont {Huberman}}]{YSWAH:2012}%
  \BibitemOpen
  \bibfield  {author} {\bibinfo {author} {\bibfnamefont {M.}~\bibnamefont
  {Ye}}, \bibinfo {author} {\bibfnamefont {T.}~\bibnamefont {Sandholm}},
  \bibinfo {author} {\bibfnamefont {C.}~\bibnamefont {Wang}}, \bibinfo {author}
  {\bibfnamefont {C.}~\bibnamefont {Aperjis}}, \ and\ \bibinfo {author}
  {\bibfnamefont {B.~A.}\ \bibnamefont {Huberman}},\ }\bibfield  {title}
  {\enquote {\bibinfo {title} {Collective attention and the dynamics of group
  deals},}\ }in\ \href@noop {} {\emph {\bibinfo {booktitle} {Proc. 21st Int.
  Conf. WWW}}}\ (\bibinfo {organization} {ACM},\ \bibinfo {year} {2012})\ pp.\
  \bibinfo {pages} {1205--1212}\BibitemShut {NoStop}%
\bibitem [{\citenamefont {Brin}\ and\ \citenamefont {Page}(1998)}]{BP:1998}%
  \BibitemOpen
  \bibfield  {author} {\bibinfo {author} {\bibfnamefont {S.}~\bibnamefont
  {Brin}}\ and\ \bibinfo {author} {\bibfnamefont {L.}~\bibnamefont {Page}},\
  }\bibfield  {title} {\enquote {\bibinfo {title} {The anatomy of a large-scale
  hypertextual {Web} search engine},}\ }\href@noop {} {\bibfield  {journal}
  {\bibinfo  {journal} {Comp. Net. ISDN}\ }\textbf {\bibinfo {volume} {30}},\
  \bibinfo {pages} {107--117} (\bibinfo {year} {1998})}\BibitemShut {NoStop}%
\bibitem [{\citenamefont {Craswell}\ and\ \citenamefont
  {Szummer}(2007)}]{CS:2007}%
  \BibitemOpen
  \bibfield  {author} {\bibinfo {author} {\bibfnamefont {N.}~\bibnamefont
  {Craswell}}\ and\ \bibinfo {author} {\bibfnamefont {M.}~\bibnamefont
  {Szummer}},\ }\bibfield  {title} {\enquote {\bibinfo {title} {Random walks on
  the click graph},}\ }in\ \href@noop {} {\emph {\bibinfo {booktitle} {Proc.
  30th Annual Int. ACM SIGIR CRDIR}}}\ (\bibinfo {organization} {ACM},\
  \bibinfo {year} {2007})\ pp.\ \bibinfo {pages} {239--246}\BibitemShut
  {NoStop}%
\bibitem [{\citenamefont {Fagin}\ \emph {et~al.}(2001)\citenamefont {Fagin},
  \citenamefont {Karlin}, \citenamefont {Kleinberg}, \citenamefont {Raghavan},
  \citenamefont {Rajagopalan}, \citenamefont {Rubinfeld}, \citenamefont
  {Sudan},\ and\ \citenamefont {Tomkins}}]{Faginetal:2001}%
  \BibitemOpen
  \bibfield  {author} {\bibinfo {author} {\bibfnamefont {R.}~\bibnamefont
  {Fagin}}, \bibinfo {author} {\bibfnamefont {A.~R.}\ \bibnamefont {Karlin}},
  \bibinfo {author} {\bibfnamefont {J.}~\bibnamefont {Kleinberg}}, \bibinfo
  {author} {\bibfnamefont {P.}~\bibnamefont {Raghavan}}, \bibinfo {author}
  {\bibfnamefont {S.}~\bibnamefont {Rajagopalan}}, \bibinfo {author}
  {\bibfnamefont {R.}~\bibnamefont {Rubinfeld}}, \bibinfo {author}
  {\bibfnamefont {M.}~\bibnamefont {Sudan}}, \ and\ \bibinfo {author}
  {\bibfnamefont {A.}~\bibnamefont {Tomkins}},\ }\bibfield  {title} {\enquote
  {\bibinfo {title} {Random walks with ``back buttons''},}\ }\href@noop {}
  {\bibfield  {journal} {\bibinfo  {journal} {Ann. Appl. Prob.}\ }\textbf
  {\bibinfo {volume} {11}},\ \bibinfo {pages} {810--862} (\bibinfo {year}
  {2001})}\BibitemShut {NoStop}%
\bibitem [{\citenamefont {Meiss}\ \emph {et~al.}(2010)\citenamefont {Meiss},
  \citenamefont {Gon{\c{c}}alves}, \citenamefont {Ramasco}, \citenamefont
  {Flammini},\ and\ \citenamefont {Menczer}}]{MGRFM:2010}%
  \BibitemOpen
  \bibfield  {author} {\bibinfo {author} {\bibfnamefont {M.~R.}\ \bibnamefont
  {Meiss}}, \bibinfo {author} {\bibfnamefont {B.}~\bibnamefont
  {Gon{\c{c}}alves}}, \bibinfo {author} {\bibfnamefont {J.~J.}\ \bibnamefont
  {Ramasco}}, \bibinfo {author} {\bibfnamefont {A.}~\bibnamefont {Flammini}}, \
  and\ \bibinfo {author} {\bibfnamefont {F.}~\bibnamefont {Menczer}},\
  }\bibfield  {title} {\enquote {\bibinfo {title} {Agents, bookmarks and
  clicks: a topical model of web navigation},}\ }in\ \href@noop {} {\emph
  {\bibinfo {booktitle} {Proc. 21st ACM CHH}}}\ (\bibinfo {organization}
  {ACM},\ \bibinfo {year} {2010})\ pp.\ \bibinfo {pages} {229--234}\BibitemShut
  {NoStop}%
\bibitem [{\citenamefont {Chierichetti}\ \emph {et~al.}(2012)\citenamefont
  {Chierichetti}, \citenamefont {Kumar}, \citenamefont {Raghavan},\ and\
  \citenamefont {Sarl{\'o}s}}]{CKRS:2012}%
  \BibitemOpen
  \bibfield  {author} {\bibinfo {author} {\bibfnamefont {F.}~\bibnamefont
  {Chierichetti}}, \bibinfo {author} {\bibfnamefont {R.}~\bibnamefont {Kumar}},
  \bibinfo {author} {\bibfnamefont {P.}~\bibnamefont {Raghavan}}, \ and\
  \bibinfo {author} {\bibfnamefont {T.}~\bibnamefont {Sarl{\'o}s}},\ }\bibfield
   {title} {\enquote {\bibinfo {title} {Are {Web} users really {Markovian}?}}\
  }in\ \href@noop {} {\emph {\bibinfo {booktitle} {Proc. 21st Int. Conf.
  WWW}}}\ (\bibinfo {organization} {ACM},\ \bibinfo {year} {2012})\ pp.\
  \bibinfo {pages} {609--618}\BibitemShut {NoStop}%
\bibitem [{\citenamefont {Tang}, \citenamefont {Wang},\ and\ \citenamefont
  {Liu}(2012)}]{TWL:2012}%
  \BibitemOpen
  \bibfield  {author} {\bibinfo {author} {\bibfnamefont {L.}~\bibnamefont
  {Tang}}, \bibinfo {author} {\bibfnamefont {X.}~\bibnamefont {Wang}}, \ and\
  \bibinfo {author} {\bibfnamefont {H.}~\bibnamefont {Liu}},\ }\bibfield
  {title} {\enquote {\bibinfo {title} {Community detection via heterogeneous
  interaction analysis},}\ }\href@noop {} {\bibfield  {journal} {\bibinfo
  {journal} {Data Min. Knowl. Disc.}\ }\textbf {\bibinfo {volume} {25}},\
  \bibinfo {pages} {1--33} (\bibinfo {year} {2012})}\BibitemShut {NoStop}%
\bibitem [{\citenamefont {Kuhn}, \citenamefont {Perc},\ and\ \citenamefont
  {Helbing}(2014)}]{KPH:2014}%
  \BibitemOpen
  \bibfield  {author} {\bibinfo {author} {\bibfnamefont {T.}~\bibnamefont
  {Kuhn}}, \bibinfo {author} {\bibfnamefont {M.}~\bibnamefont {Perc}}, \ and\
  \bibinfo {author} {\bibfnamefont {D.}~\bibnamefont {Helbing}},\ }\bibfield
  {title} {\enquote {\bibinfo {title} {Inheritance patterns in citation
  networks reveal scientific memes},}\ }\href@noop {} {\bibfield  {journal}
  {\bibinfo  {journal} {Phys. Rev. X}\ }\textbf {\bibinfo {volume} {4}},\
  \bibinfo {pages} {041036} (\bibinfo {year} {2014})}\BibitemShut {NoStop}%
\bibitem [{\citenamefont {Sanl{\i}}\ and\ \citenamefont
  {Lambiotte}(2015)}]{SL:2015}%
  \BibitemOpen
  \bibfield  {author} {\bibinfo {author} {\bibfnamefont {C.}~\bibnamefont
  {Sanl{\i}}}\ and\ \bibinfo {author} {\bibfnamefont {R.}~\bibnamefont
  {Lambiotte}},\ }\bibfield  {title} {\enquote {\bibinfo {title} {Local
  variation of hashtag spike trains and popularity in {Twitter}},}\ }\href@noop
  {} {\bibfield  {journal} {\bibinfo  {journal} {PloS One}\ }\textbf {\bibinfo
  {volume} {10}},\ \bibinfo {pages} {e0131704} (\bibinfo {year}
  {2015})}\BibitemShut {NoStop}%
\bibitem [{\citenamefont {Zhu}\ \emph {et~al.}(2016)\citenamefont {Zhu},
  \citenamefont {Yin}, \citenamefont {Ma},\ and\ \citenamefont
  {Hu}}]{ZYMH:2016}%
  \BibitemOpen
  \bibfield  {author} {\bibinfo {author} {\bibfnamefont {H.}~\bibnamefont
  {Zhu}}, \bibinfo {author} {\bibfnamefont {X.}~\bibnamefont {Yin}}, \bibinfo
  {author} {\bibfnamefont {J.}~\bibnamefont {Ma}}, \ and\ \bibinfo {author}
  {\bibfnamefont {W.}~\bibnamefont {Hu}},\ }\bibfield  {title} {\enquote
  {\bibinfo {title} {Identifying the main paths of information diffusion in
  online social networks},}\ }\href@noop {} {\bibfield  {journal} {\bibinfo
  {journal} {Physica A}\ }\textbf {\bibinfo {volume} {452}},\ \bibinfo {pages}
  {320--328} (\bibinfo {year} {2016})}\BibitemShut {NoStop}%
\bibitem [{\citenamefont {Piedrahita}\ \emph {et~al.}(2018)\citenamefont
  {Piedrahita}, \citenamefont {Borge-Holthoefer}, \citenamefont {Moreno},\ and\
  \citenamefont {Gonzalez-Bailon}}]{PBHMGB:2018}%
  \BibitemOpen
  \bibfield  {author} {\bibinfo {author} {\bibfnamefont {P.}~\bibnamefont
  {Piedrahita}}, \bibinfo {author} {\bibfnamefont {J.}~\bibnamefont
  {Borge-Holthoefer}}, \bibinfo {author} {\bibfnamefont {Y.}~\bibnamefont
  {Moreno}}, \ and\ \bibinfo {author} {\bibfnamefont {S.}~\bibnamefont
  {Gonzalez-Bailon}},\ }\bibfield  {title} {\enquote {\bibinfo {title} {The
  contagion effects of repeated activation in social networks},}\ }\href@noop
  {} {\bibfield  {journal} {\bibinfo  {journal} {Soc. Net.}\ }\textbf {\bibinfo
  {volume} {54}},\ \bibinfo {pages} {326--335} (\bibinfo {year}
  {2018})}\BibitemShut {NoStop}%
\end{thebibliography}

%
\end{document}